 \documentclass[final,5p,times,twocolumn]{elsarticle}

\usepackage{amssymb}
\usepackage{lipsum}
\usepackage{amsmath,amssymb,amsfonts}
\usepackage{algorithmic}
\usepackage{graphicx}
\usepackage{textcomp}
\usepackage{xcolor}
\usepackage{float} %设置图片浮动位置的宏包
 \usepackage{epstopdf}
 \usepackage{multirow}
\usepackage{makecell}
\usepackage[caption=false,font=footnotesize,labelfont=rm,textfont=rm]{subfig}
\DeclareUnicodeCharacter{230A}{\lfloor}
\journal{Nuclear Physics B}

\begin{document}

\begin{frontmatter}

%% Title, authors and addresses

%% use the tnoteref command within \title for footnotes;
%% use the tnotetext command for theassociated footnote;
%% use the fnref command within \author or \address for footnotes;
%% use the fntext command for theassociated footnote;
%% use the corref command within \author for corresponding author footnotes;
%% use the cortext command for theassociated footnote;
%% use the ead command for the email address,
%% and the form \ead[url] for the home page:
%% \title{Title\tnoteref{label1}}
%% \tnotetext[label1]{}
%% \author{Name\corref{cor1}\fnref{label2}}
%% \ead{email address}
%% \ead[url]{home page}
%% \fntext[label2]{}
%% \cortext[cor1]{}
%% \affiliation{organization={},
%%             addressline={},
%%             city={},
%%             postcode={},
%%             state={},
%%             country={}}
%% \fntext[label3]{}

\title{Causal Disentanglement for Regulating Social Influence Bias in Social Recommendation}

%% use optional labels to link authors explicitly to addresses:
%% \author[label1,label2]{}
%% \affiliation[label1]{organization={},
%%             addressline={},
%%             city={},
%%             postcode={},
%%             state={},
%%             country={}}
%%
%% \affiliation[label2]{organization={},
%%             addressline={},
%%             city={},
%%             postcode={},
%%             state={},
%%             country={}}

% \author[inst1]{Author One}

% \affiliation[inst1]{organization={Department One},%Department and Organization
%             addressline={Address One}, 
%             city={City One},
%             postcode={00000}, 
%             state={State One},
%             country={Country One}}

% \author[inst2]{Author Two}
% \author[inst1,inst2]{Author Three}

% \affiliation[inst2]{organization={Department Two},%Department and Organization
%             addressline={Address Two}, 
%             city={City Two},
%             postcode={22222}, 
%             state={State Two},
%             country={Country Two}}

\author[first]{Li Wang}
\ead{li.wang-13@student.uts.edu.au}
\author[first]{Min Xu\corref{cor1}}
\ead{Min.Xu@uts.edu.au}
\author[second]{Quangui\ Zhang}
\ead{zhqgui@126.com}
\author[first]{Yunxiao\ Shi}
\ead{Yunxiao.Shi@student.uts.edu.au}
\author[first]{Qiang\ Wu}
\ead{Qiang.Wu@uts.edu.au}
\affiliation[first]{organization={School of Electrical and Data Engineering, University of Technology Sydney},%Department and Organization
            addressline={15, Broadway, Ultimo}, 
            city={Sydney},
            postcode={2000}, 
            state={NSW},
            country={Australia}}
\affiliation[second]{organization={School of Artificial Intelligence,
Chongqing University of Arts and Sciences},%Department and Organization
            addressline={319, Honghe Avenue, Yongchuan District}, 
            city={Chongqing},
            postcode={402160}, 
            country={China}}
\cortext[cor1]{Corresponding author}

\begin{abstract}
%% Text of abstract
Social recommendation systems face the problem of social influence bias, which can lead to an overemphasis on recommending items that friends have interacted with. Addressing this problem is crucial, and existing methods often rely on techniques such as weight adjustment or leveraging unbiased data to eliminate this bias.
However, we argue that not all biases are detrimental, i.e., some items recommended by friends may align with the user’s interests. Blindly eliminating such biases could undermine these positive effects, potentially diminishing recommendation accuracy.
In this paper, we propose a \textbf{C}ausal \textbf{D}isentanglement-based framework for \textbf{R}egulating \textbf{S}ocial influence \textbf{B}ias in social recommendation, named CDRSB, to improve recommendation performance. From the perspective of causal inference, we find that the user social network could be regarded as a confounder between the user and item embeddings (treatment) and ratings (outcome).  Due to the presence of this social network confounder, two paths exist from user and item embeddings to ratings: a non-causal social influence path and a causal interest path. Building upon this insight, we propose a disentangled encoder that focuses on disentangling user and item embeddings into interest and social influence embeddings. Mutual information-based objectives are designed to enhance the distinctiveness of these disentangled embeddings, eliminating redundant information.
Additionally, a regulatory decoder that employs a weight calculation module to dynamically learn the weights of social influence embeddings for effectively regulating social influence bias has been designed.
Experimental results on four large-scale real-world datasets Ciao, Epinions, Dianping, and Douban book demonstrate the effectiveness of CDRSB compared to state-of-the-art baselines. 
% We release our code at https://github.com/Lili1013/CDRSB.
\end{abstract}

% %%Graphical abstract
% \begin{graphicalabstract}
% \includegraphics{grabs}
% \end{graphicalabstract}

% %%Research highlights
% \begin{highlights}
% \item Research highlight 1
% \item Research highlight 2
% \end{highlights}


\end{frontmatter}

%% \linenumbers

%% main text
\section{Introduction}
\label{introduction}

Social recommendation systems (SR) play a crucial role in addressing the data-sparsity challenge and enhancing recommendation performance by incorporating social network information \cite{fan2018deep,wu2019neural,zhu2022si}. 
These systems leverage data from users' social interactions, encompassing friendships, shared content, comments and likes on social networks, to learn comprehensive and personalized user and item representations.
Notable methods like SocialMF \cite{jamali2010matrix}, GraphRec \cite{fan2019graph}, and DiffNet \cite{wu2019neural} have introduced innovative approaches, such as integrating social network information into traditional matrix factorization methods or utilizing Graph Neural Networks (GNN) to model high-order social relationships.

Despite these advancements, a critical limitation arises due to the potential presence of social influence bias in the acquired user and item representations, which may not accurately reflect users' genuine interests. 
This bias originates from the phenomenon where individuals, under the influence of their social circles, might make choices that deviate from their personal preferences.
In recent years, some existing methods \cite{sheth2023causal,li2023causal,sheth2022causal} have focused on mitigating social influence bias generated from network information.
For example, SIDR \cite{sheth2023causal} disentangles user and item representations into three latent factors: user interest, item popularity, and user social influence, and then mitigates the social influence bias.
DENC \cite{li2023causal} proposes an exposure model and a deconfounding model to effectively control and eliminate social influence bias.
Conversely, D2Rec \cite{sheth2022causal} utilizes network information to disentangle user and item representations into exposure, confounder, and prediction factors.
It designs a reweighting function to mitigate social influence bias. While SIDR \cite{sheth2023causal} employs causal disentanglement to separate social influence, its primary emphasis is on mitigating social influence bias rather than strategically leveraging it.

Rather than indiscriminately mitigating social influence bias, we argue that it is crucial to recognize that not all biases are harmful.
Some users perceive recommendations from friends as thoughtful selections, indicating high quality and alignment with their interests.
In such cases, the influence exerted by friends has a positive impact on users, and these products are worth recommending. Blindly mitigating this bias may lead to the loss of essential information, preventing the recommendations that align with users' interests. Thus, a dilemma arises: eliminating social influence bias sacrifices meaningful recommendations, while preserving it may lead to undesirable social conformity.
Therefore, it is crucial and urgent to propose a method that can reasonably regulate social influence bias to enhance recommendation performance. Such a method should preserve positive social influence bias while mitigating its negative counterpart.

\begin{figure*}[!t]
\centering
\subfloat[Fork structure of causal graph.] {\includegraphics[width=0.18\textwidth,height=1.7in]{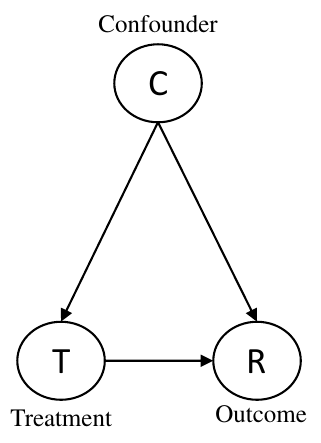}
\label{causal_graph}
}
\hfil
\subfloat[Causal graph of social recommendations.]{\includegraphics[width=0.25\textwidth,height=1.7in]{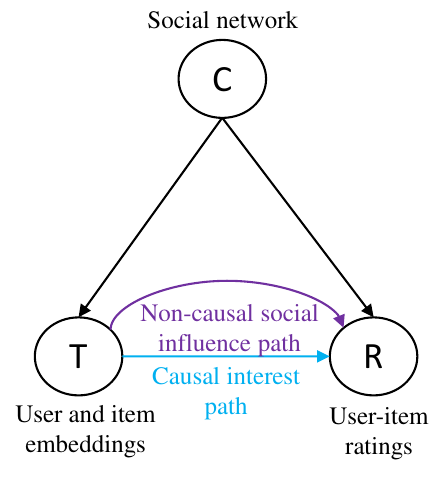}
\label{causal_graph_SR}
}
\hfil
\subfloat[Causal graph of CDRSB.]{\includegraphics[width=0.3\textwidth,height=1.7in]{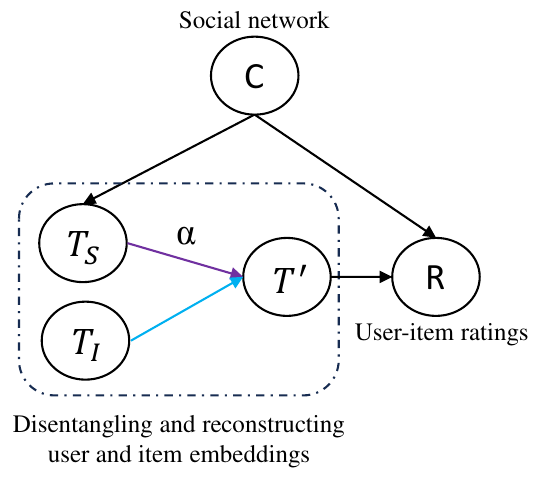}
\label{causal_graph_CDRSB}
}
\hfil
\caption{(a) Fork structure of causal graph: confounder affects both the treatment and the outcome. (b) In social recommendations, we treat the social network as a confounder, with user and item embeddings as the treatment and user-item ratings as the outcome. Due to the existence of the social network confounder, there are two paths between user and item embeddings and user-item ratings: a non-causal social influence path and a causal interest path. (c) CDRSB disentangles the user and item embeddings into social influence and interest components and learns dynamic weights of social influence embeddings to fuse them, thereby effectively regulating social influence bias. $C$: social network, $T$: user and item embeddings, $R$: user-item ratings, $T_S$: social influence embeddings, $T_I$: interest embeddings, $T'$: reconstructed user and item representations, $\alpha$: the weight of social influence embeddings.}
\label{motovation}
\end{figure*}

To gain a deeper understanding of how social influence affects recommendations, we introduce the causal graph to analyze this process.
Motivated by the fork structure of the causal model proposed by Pearl \cite{pearl2000models}, as shown in Figure \ref{causal_graph}, 
the confounder is a variable that is related to both the treatment variable (the cause) and the outcome variable (the effect). The presence of a confounder can lead to a spurious correlation between the treatment and outcome variables, making it challenging to establish the causal relationship.
In the context of social recommendations, as illustrated in Figure \ref{causal_graph_SR}, we can regard user and item embeddings as the treatment and user-item ratings as the outcome. The social network emerges as a potential confounder since it exerts simultaneous influence on both the user and item embeddings as well as user-item ratings.
Due to the existence of the social network confounder, there are two paths from the user and item embeddings (treatment) $T$ and the user-item ratings (outcome) $R$, including the non-causal social influence path ($C\to T\to R$) introduced by the confounder and the causal interest ($T\to R$) path which represents the reason why a user likes an item.
We refer to the bias introduced by the social influence path as social influence bias.
This bias has the undesirable effect of bias amplification because it increases the exposure probability of items that friends interact with, even if these items do not match the user's interests. However, some of the items recommended by friends may align with the user's interests and deserve to be recommended.
Therefore, it's crucial to design a method that can effectively regulate social influence bias, preserving its positive effects while mitigating the negative ones.

Based on the above analysis, we propose a causal disentanglement-based framework for regulating social influence bias in social recommendations, named CDRSB, as illustrated in Figure \ref{causal_graph_CDRSB}.
We assume that the treatment (user and item embeddings) can be causally decomposed into two independent embeddings: interest embedding which represents the user's real preferences, and social influence embedding which indicates the social influence bias.
We design a disentangled encoder aimed at separating interest and social influence embeddings, along with a regulatory decoder that designs a weight calculation module to reasonably regulate social influence bias.
Specifically, for the disentangled encoder, 
we first learn user and item embeddings via GNN-based learning networks with the user social network and user-item interaction network.
Then, we design a causal disentanglement component to separate the user and item embeddings into interest and social influence embeddings.
To make these two components independent of each other and contain more semantic information, we introduce mutual information-based objectives.
Regarding the regulatory decoder, we first introduce a weight calculation module to learn varying weights of social influence embeddings, which could regulate the social influence bias. These weights are then utilized to fuse interest and social influence embeddings to learn more accurate user and item representations.
The main contributions of this work are as follows:
\begin{itemize}
\item Based on the non-causal social influence path and the causal interest path introduced by the social network confounder, we propose a disentangled encoder. This encoder disentangles user and item embeddings into interest and social influence embeddings. Mutual information-based objectives are designed to ensure the separation of these disentangled embeddings.
\item We propose a regulatory decoder that introduces a weight calculation module to regulate social influence bias and learn more accurate user and item representations, enhancing the model's performance.
\item Extensive experiments are conducted on four large-scale real-world datasets Ciao, Epinions, Dianping, and Douban book.
The comprehensive results demonstrate the effectiveness of CDRSB compared to several state-of-the-art baselines.
\end{itemize}
The remainder of this work is organized as follows: In Section 2, we provide an overview of the relevant work. Section 3 presents the detailed methodology of the CDRSB. 
We then conduct extensive experiments on four public datasets and compare the results with baselines in Section 4.
Finally, in Section 5, we provide a conclusion summarizing our work and outline future research directions.

\section{Related Work}
In this section, we present several relevant studies related to social recommendation, disentangled representation learning in recommendation, and causal recommendation.

\noindent \textbf{Social Recommendation}. In recent years, social recommendation systems have gained significant attention due to the widespread adoption of social media platforms and the increasing number of users.
The social network provides rich social relationships and interaction information among users, which can solve the long-standing data-sparsity problem \cite{ma2008sorec,wang2014hgmf,chen2019samwalker}.
%In addition, it can be leveraged to gain a deeper understanding of users' interests, preferences and behavior patterns.
Existing social recommendations could be categorized as matrix factorization (MF)-based \cite{jamali2010matrix,ma2011recommender} and graph neural network (GNN)-based approaches \cite{wu2019neural,fan2019graph,yang2021consisrec,wu2020diffnet++}.
MF-based methods usually jointly factorize the user-item interaction matrix and user-social relationship matrix or add a regularizer to restrain user and item embeddings.
SocialMF \cite{jamali2010matrix} %[A Matrix Factorization Technique with Trust Propagation for Recommendation in Social Networks]
incorporates the user's social network into traditional collaborative filtering models for recommendation.
In \cite{ma2011recommender}, authors introduce a social regularization term to incorporate users' social relationships into the recommendation model. 
In contrast, GNN-based methods utilize the connectivity of graphs to directly model user and item embeddings.
GraphRec \cite{fan2019graph} %[graph neural networks for social recommendation] 
introduce a graph attention mechanism with both the user social network and the user-item interaction graph to learn user and item embeddings.
ConsisRec \cite{yang2021consisrec} %[ConsisRec: Enhancing GNN for Social Recommendation via Consistent Neighbor Aggregation]
is an improved version of GraphRec that addresses the social inconsistency problem.
To improve the performance of social recommendation, DiffNet \cite{wu2019neural} and DiffNet++ \cite{wu2020diffnet++} learn the social influence diffusion process.
However, these methods ignore the problem of social influence bias, which may degrade the recommendation performance.

\noindent \textbf{Disentangled Representation Learning in Recommendation}.
Disentangled representation learning has been recognized as an effective way to enhance the robustness and interpretability of models \cite{wang2022disentangled,wu2020improving}.
Disentangled representation learning has been applied in generative recommendations \cite{ma2019learning,burgess2018understanding,bouchacourt2018multi} and graph recommendations \cite{wang2020disentangled,wang2020disenhan,li2022disentangled}.
Authors in \cite{ma2019learning} % [Learning Disentangled Representations for Recommendation] 
propose a model called MacridVAE, which is a disentangled variational auto-encoder capable of learning representations from user behavior.
This model achieves both macro-disentanglement of high-level concepts and micro-disentanglement of isolated low-level factors.
DGCF \cite{wang2020disentangled} learns disentangled representations that capture fine-grained user intents from the user-item interaction graph.
DisenHAN \cite{wang2020disenhan} learns disentangled user/item representations from various aspects in a heterogeneous information network, utilizing meta relations to decompose high-order connectivity between node pairs. 
Recently, disentangled representation learning has been applied to causal recommendation systems \cite{zheng2021disentangling,liu2021mitigating,si2022model}. 
DICE \cite{zheng2021disentangling} constructs cause-specific data according to causal effect and disentangles user and item embeddings into interest and conformity components.
DIB \cite{liu2021mitigating} % [mitigating confounding bias in recommendation via information bottleneck] 
mitigates confounding bias by decomposing user and item embeddings into unbiased and biased components via information bottleneck.
Authors in \cite{si2022model} % [A model-agnostic causal embedding learning framework for recommendation using search data] 
utilize user search data to decouple corresponding actual preferences, 
providing a model-agnostic approach to causal embedding learning in recommendation systems.
Nevertheless, these methods may not be suitable for social recommendations where social influence plays an important role in modeling user preferences.

\noindent \textbf{Causal Recommendation}.
In contrast to traditional recommendation systems that predominantly emphasize correlational patterns, causal recommendation systems delve into the realm of causality. Their objective is to discern and address the causal relationships between user actions and the recommended outcomes \cite{gao2022causal}. 
These systems leverage methodologies rooted in causal inference, such as inverse propensity weighting \cite{li2023causal,zhang2020large}, backdoor adjustment \cite{he2023addressing,wang2021deconfounded}, frontdoor adjustment \cite{xu2023deconfounded,zhu2022mitigating}, and counterfactual inference \cite{wei2021model,he2022mitigating}, to understand and mitigate various biases like confounding bias, selection bias or spurious correlations.
For example, AutoDebias \cite{chen2021autodebias} leverages uniform data collected by a random logging policy and meta-learning technique to mitigate various biases.
Zhang et al. \cite{zhang2020large} introduce the Multi-IPW model, employing a multi-task learning approach to estimate Inverse Propensity Scores (IPS) and simultaneously mitigate selection bias.
DCR \cite{he2023addressing} introduces the notion of item confounding features and employs backdoor adjustment combined with a mixture-of-experts (MoE) strategy to alleviate the spurious correlations arising from them.
DCCF \cite{xu2023deconfounded} utilizes frontdoor adjustment to alleviate the confounding bias.
Wei et al. \cite{wei2021model} focus specifically on countering popularity bias through the application of counterfactual inference.
Recently, some causal disentanglement-based methods have been proposed to achieve unbiased recommendations \cite{li2023causal,sheth2022causal}.
For example, DENC \cite{li2023causal} and D2Rec \cite{sheth2022causal} both focus on disentangling three causes: inherent, confounder, and exposure factors, and mitigating bias produced by network information.
Nevertheless, these approaches mainly eliminate biases, it is worth noting that certain biases, such as popularity bias and social influence bias, can occasionally be beneficial for learning user preferences \cite{chen2023bias}.
\section{Methodology}
In this part, we first present the definitions and notations used in this paper and then provide a concise overview of the overall framework. Following that, we introduce each component in detail.
\subsection{Definitions and Notations}
Let $U=\{ u_1, u_2,...,u_n \}$ and $V=\{ v_1, v_2,...,v_m \}$ denote the sets of users and items, where $n$ and $m$ are the number of users and items, respectively.
$\textbf{p}_i\in \mathbb{R}^d$ represents the ID embedding of user $u_i$ and $\textbf{q}_j \in \mathbb{R}^d$ denotes the ID embedding of item $v_j$, $d$ is the embedding size.
$\bf{Y} \in \mathbb{R}^{m\times n}$ represents the user-item rating matrix, where $y_{ij}\in \textbf{Y}$ denotes the rating given by user $u_i$ to item $v_j$. Let use $\bf{T}\in \mathbb{R}^{n\times n}$ to denote the user social network,
where $T_{ij}=1$ if there is a relation between $u_i$ and $u_j$.
%Similarly, $\bf{S}\in \mathbb{R}^{m\times m}$ is the item relation network.
$\textbf{e}_{ij}^u$ represents the rating embedding of user $u_i$ on item $v_j$ that $u_i$ has interacted with, and $\textbf{e}_{ji}^v$ denotes the rating embedding of item $v_j$ rated by user $u_i$.
We use $\textbf{z}_i^u$, $\textbf{c}_i^u$, $\textbf{z}_j^v$, and $\textbf{c}_j^v$ to denote the interest embedding and social influence embedding of user $u_i$ and item $v_j$, respectively. 
$\textbf{x}_i^u$ and $\textbf{x}_j^v$ represent the embeddings learned by a GNN-based network of user $u_i$ and item $v_j$.
In addition, $\textbf{h}_i^u$ and $\textbf{h}_j^v$ denote the reconstructed user and item representations.
The mathematical notations used in this paper are summarized in Table \ref{notations}.
\begin{table}[h]
\centering
\caption{Notations.}\label{notations}
\begin{tabular}{c|c}
\hline
Symbols & Definitions and Notations  \\\hline
$\textbf{Y}$ & user-item rating matrix\\
$U$ & user set\\
$V$ & item set\\
$\textbf{T}$ & user social network\\
% $\textbf{S}$ & item relation network\\
$C(i)$ & the set of items which user $u_i$ have interacted with\\
$N(i)$ & the set of neighbouring users connected to user $u_i$\\
$B(j)$ & the set of users who interact with item $v_j$\\
%$A(j)$ & the set of items which are similar to item $v_j$\\
$\textbf{p}_i$ & ID embedding of user $u_i$\\
$\textbf{q}_j$ & ID embedding of item $v_j$\\
$\textbf{x}_i^u$ & \makecell{embedding learned from \\a GNN-based network of user $u_i$}\\
$\textbf{x}_j^v$ & \makecell{embedding learned from \\a GNN-based network of item $v_j$}\\
$\textbf{z}_i^u$ &   interest embedding of user $u_i$\\
$\textbf{c}_i^u$ &   social influence embedding of user $u_i$\\
$\textbf{z}_j^v$ &   interest embedding of item $v_j$\\
$\textbf{c}_j^v$ &   social influence embedding of item $v_j$\\
$\textbf{e}_{ij}^u$ & \makecell{rating embedding of user $u_i$ \\for item $v_j$ that $u_i$ has interacted with}\\
$\textbf{e}_{ji}^v$ & rating embedding of item $v_j$ by user $u_i$\\
$\textbf{r}_{ij}^v$ & item embedding from item set $C(i)$ of user $u_i$\\
$\textbf{r}_{ji}^u$ & user embedding from user set $B(j)$ of item $v_j$\\
$\textbf{t}_{ij}^u$ & user embedding from user set $N(i)$ of user $u_i$\\
$\textbf{h}_i^u$ & reconstructed representation of user $u_i$\\
$\textbf{h}_j^v$ & reconstructed representation of item $v_j$\\
\hline
\end{tabular}
\end{table}

\subsection{An overview of the proposed model}
\begin{figure*}[h] %H为当前位置，!htb为忽略美学标准，htbp为浮动图形
\centering %图片居中
\includegraphics[width=1.0\textwidth]{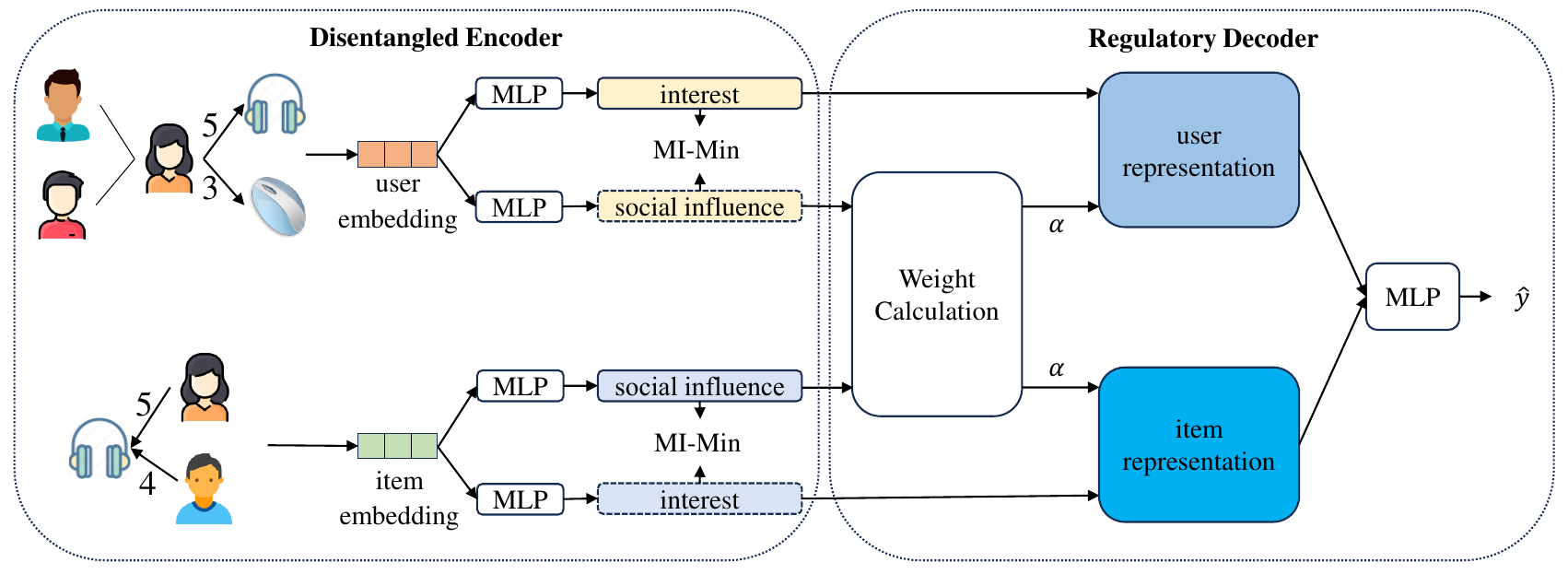} %插入图片，[]中设置图片大小，{}中是图片文件名
\caption{The overall framework of CDRSB. It contains two modules: (1) a disentangled encoder that disentangles user and item embeddings learned from a GNN-based network into interest and social influence embeddings. We minimize mutual information-based objectives to reduce redundancy and ensure the separation of these disentangled embeddings. (2) a regulatory decoder that learns dynamic weights to combine interest and social influence embeddings into final user and item representations, achieving reasonable utilization of social influence bias.} %最终文档中希望显示的图片标题
\label{fig2.framework} %用于文内引用的标签
\end{figure*}

We propose a causal disentanglement-based framework for regulating social influence bias in social recommendation, named CDRSB. Figure \ref{fig2.framework} shows the overall architecture of the model.

This framework mainly includes two modules:
(1) \textbf{Disentangled Encoder}. We aim to disentangle user and item embeddings learned from network information into interest and social influence embeddings.
(2) \textbf{Regulatory Decoder}. We introduce a weight calculation module to regulate the social influence bias, which could improve the recommendation performance.
\subsection{Disentangled Encoder}
In this section, based on the non-causal social influence path and causal interest path of the causal graph for social recommendations, our focus is on disentangling user and item embeddings into interest and social influence embeddings. 
Initially, we employ a GNN-based learning network with the user-item interaction graph and user-social graph to learn user and item embeddings.
Subsequently, we introduce independent MLP layers to disentangle interest and social influence embeddings.
\subsubsection{GNN-based Learning Network}
 Motivated by the approach introduced in \cite{fan2019graph}, % [Graphrec], 
we design a similar GNN-based method to learn the initial user and item embeddings, which contains two layers: the embedding layer and the concatenation layer.

 \noindent \textbf{Embedding layer}: We input the user social network and user-item interaction network into the embedding layer to learn the ID embedding, neighboring item embedding, and neighboring user embedding of user $u_i$.
In addition, we also consider each user-item rating in the user-item interaction network, which represents the user's preference level.
\begin{equation}
 \begin{aligned}
 &\textbf{p}_i = \delta(\textbf{W}\cdot \textbf{o}_i+\textbf{b}), \\
&\textbf{r}_{ij, \forall j \in C(i)}^v = \delta(\textbf{W}\cdot \textbf{o}_{ij}^r+\textbf{b}),\\
&\textbf{e}_{ij, \forall j \in C(i)}^u = \delta(\textbf{W}\cdot \textbf{o}_{ij}^e+\textbf{b}),\\
 &\textbf{t}_{ij,\forall j \in N(i)}^u = \delta(\textbf{W}\cdot \textbf{o}_{ij}^t+\textbf{b}),
 \end{aligned}
 \end{equation}
 
where $\textbf{p}_i$ and $\textbf{o}_i$ represent the ID embedding and one-hot vectors of user $u_i$.
$\textbf{r}_{ij}^v$ and $\textbf{o}_{ij}^r$ are the embedding and one-hot vectors of item $v_j$ that user $u_i$ has interacted with.
$\textbf{e}_{ij}^u$ and $\textbf{o}_{ij}^e$ are the rating embedding and one-hot vectors of user $u_i$ for the item $v_j$ that user $u_i$ has interacted with.
$\textbf{t}_{ij}^u$ and $\textbf{o}_{ij}^t$ are the embedding and one-hot vectors of user $u_j$ that user $u_i$ has connected with.
$\delta$ denotes the non-linear activation function, $\textbf{W}$ and $\textbf{b}$ represent the weight matrix and bias vector, respectively.
 
\noindent \textbf{Concatenation layer}: In this part, we concatenate user ID embedding $\textbf{p}_i$, neighboring item embedding $\textbf{r}_{ij}^v$, and rating embedding $\textbf{e}_{ij}^u$ from the user-item interaction network, as well as the neighboring user embedding $\textbf{t}_{ij}^u$ from the user-social network to learn the user embedding $\textbf{x}_i^u$.
  \begin{equation}
 \begin{split}
 \textbf{x}_i^u = \delta(\textbf{W}\cdot (\textbf{p}_i&\oplus Agg_{items}({\textbf{r}_{ij}^v, \forall j \in C(i)}) \\
 &\oplus Agg_{ratings}({\textbf{e}_{ij}^u, \forall j \in C(i)}) \\
 &\oplus Agg_{users}({\textbf{t}_{ij}^u, \forall j \in N(i)}))+\textbf{b}),
 \end{split}
 \end{equation}

 % \begin{equation}
 % \begin{split}
 % \textbf{r}_i^u = \delta(\textbf{W}\cdot (\textbf{p}_i\oplus Agg_{items}({\textbf{r}_{ij}^v, \forall j \in C(i)})\oplus \\
 % Agg_{ratings}({\textbf{e}_{ij}^u, \forall j \in C(i)}))+\textbf{b}),
 % \end{split}
 % \end{equation}
 % \begin{equation}
 % \textbf{t}_i^u = \delta(\textbf{W}\cdot (\textbf{p}_i\oplus Agg_{users}({\textbf{t}_{ij}^u, \forall j \in N(i)}))+\textbf{b}),
 % \end{equation}
where $Agg_{items}$ represents the aggregation operation for items, $Agg_{ratings}$ represents the aggregation operation for user-item ratings, and $Agg_{users}$ represents the user aggregation function.
We have tried various aggregation methods, including sum aggregation, mean aggregation, and neural network aggregation.
Among them, the mean aggregation obtained the best results.
$C(i)$ is a set of items that user $u_i$ has interacted with.
$N(i)$ is the set of neighboring users trusted by user $u_i$. $\oplus$ represents the concatenation operation.

 % Finally, we concatenate the embeddings $\textbf{r}_i^u$ and $\textbf{t}_i^u$ to learn the user representation $\textbf{x}_i^u$,
 % \begin{equation}
 % \textbf{x}_i^u = \delta(\textbf{W}\cdot (\textbf{r}_i^u\oplus \textbf{t}_i^u)+\textbf{b}),
 % \end{equation}
 % where $\textbf{r}_i^u$ is learned from the user-item rating network and  $\textbf{t}_i^u$ is learned from the user social network.
 
Similarly, we utilize the user-item interaction network to learn the item embedding $\textbf{x}_j^v$.
  \begin{equation}
 \begin{aligned}
 &\textbf{q}_j = \delta(\textbf{W}\cdot \textbf{o}_j+\textbf{b}), \\
&\textbf{r}_{ji, \forall i \in B(j)}^u = \delta(\textbf{W}\cdot \textbf{o}_{ji}^r+\textbf{b}),\\
&\textbf{e}_{ji, \forall i \in B(j)}^v = \delta(\textbf{W}\cdot \textbf{o}_{ji}^e+\textbf{b}),\\
% &\textbf{s}_{ji,\forall i \in A(j)}^v = \delta(\textbf{W}\cdot \textbf{o}_{ji}^s+\textbf{b}),
 \end{aligned}
 \end{equation}
 \begin{equation}
 \begin{split}
 \textbf{x}_j^v = \delta(\textbf{W}\cdot (\textbf{q}_j\oplus Agg_{users}({\textbf{r}_{ji}^u, \forall i \in B(j)})\oplus\\
 Agg_{ratings}({\textbf{e}_{ji}^v, \forall i \in B(j)}))+\textbf{b}),
 \end{split}
 \end{equation}
%   \begin{equation}
% \textbf{s}_j^v = \delta(\textbf{W}\cdot (\textbf{q}_j\oplus Agg_{items}({\textbf{s}_{ji}^v, \forall i \in A(j)}))+\textbf{b}),
% \end{equation}
where $B(j)$ is the set of users who interacted with item $v_j$.
%  and $A(j)$ is the set of items that is similar to item $v_j$.
 
 %  Finally, the item representation is learned as follows,
 %  \begin{equation}
 % \textbf{x}_j^v = \delta(\textbf{W}\cdot \textbf{r}_j^v+\textbf{b}),
 % \end{equation}
 % where $\textbf{r}_j^v$ is learned from the user-item rating network. 
% and  $\textbf{s}_j^v$ is learned from the item relation network.

\subsubsection{Causal Disentanglement}
In this subsection, we design four independent MLP networks to decompose user and item embeddings into interest and social influence embeddings.
Specifically, for the user embedding $\textbf{x}_i^u$, we extract two independent interest and social influence components.
We feed $\textbf{x}_i^u$ into two separate MLP layers,

\begin{equation}
\textbf{z}_i^u=MLP(\textbf{x}_i^u;\Theta_0);  \qquad  \textbf{c}_i^u=MLP(\textbf{x}_i^u;\Theta_1).
\end{equation}

Similarly, the decomposing process of item embedding $\textbf{x}_j^v$ is as follows,

\begin{equation}
\textbf{z}_j^v=MLP(\textbf{x}_j^v;\Theta_2);  \qquad  \textbf{c}_j^v=MLP(\textbf{x}_j^v;\Theta_3),
\end{equation}
where $\Theta_0, \Theta_1, \Theta_2$ and $\Theta_3$ refer to all the parameters for the respective MLP layers.

Despite using separate networks to learn interest and social influence embeddings, we cannot guarantee the absence of redundant information. 
Therefore, we aim to minimize the mutual information between them to ensure their independence.
Mutual information is a statistical measure that quantifies the amount of information one random variable contains about another, indicating the level of dependence between these two variables.
High mutual information indicates a strong relationship or dependency between the variables, while low mutual information suggests independence or little shared information.

Traditionally, mutual information is calculated based on examples $a_i$ and $b_i$ sampled from two distributions $A$ and $B$. 
However, in our case, the true distributions of $A$ and $B$ are unknown.
Taking inspiration from recent advancements in contrastive learning and sample-based mutual information estimation, such as the CLUB framework \cite{cheng2020club}, %(Contrastive Log-ratio Upper Bound of Mutual Information), 
we estimate mutual information using the difference in conditional probabilities between positive and negative sample pairs.
We regard a sample pair with the same index as a positive pair, such as $(a_i, b_i)$, and consider a sample pair with a different index as a negative pair, such as $(a_i,b_j)$.

Since we cannot directly compute the conditional distribution $P(\textbf{c}_{i}^u|\textbf{z}_{i}^u)$ and $P(\textbf{c}_{j}^v|\textbf{z}_{j}^v)$, 
we use variational distributions $Q_{\theta_{u}}(\textbf{c}_{i}^u|\textbf{z}_{i}^u)$ and $Q_{\theta_{v}}(\textbf{c}_{j}^v|\textbf{z}_{j}^v)$ with parameters $\theta_u$ and $\theta_v$, which could be implemented by neural networks, to approximate $P(\textbf{c}_{i}^u|\textbf{z}_{i}^u)$ and $P(\textbf{c}_{j}^v|\textbf{z}_{j}^v)$. 
The output of $Q_{\theta_{u}}(\textbf{c}_{i}^u|\textbf{z}_{i}^u)$/$Q_{\theta_{v}}(\textbf{c}_{j}^v|\textbf{z}_{j}^v)$ are mean $\mu_i^u$/$\mu_j^v$ and log variance $log({\delta_i^u}^2)$/$log({\delta_j^v}^2)$.
To update the parameter $\theta_{u}$ and $\theta_{v}$, we maximize the corresponding conditional log-likelihood loss function,
 \begin{equation}
 \begin{split}
 L_{u}^{LLD}&=\frac{1}{N}\sum_{i=1}^NlogQ_{\theta_{u}}(\textbf{c}_{i}^u|\textbf{z}_{i}^u)\\
            &=\frac{1}{N}\sum_{i=1}^N-\frac{(\mu_i^u-\textbf{c}_{i}^u)^2}{e^{log({\delta_i^u}^2)}};\\
 L_{v}^{LLD}&=\frac{1}{N}\sum_{j=1}^NlogQ_{\theta_{v}}(\textbf{c}_{j}^v|\textbf{z}_{j}^v)\\
            &=\frac{1}{N}\sum_{i=1}^N-\frac{(\mu_j^v-\textbf{c}_{j}^v)^2}{e^{log({\delta_j^v}^2)}},
 \end{split}
 \end{equation}
 
 where $N$ is the number of samples. $\textbf{z}_{i}^u$, $\textbf{c}_{i}^u$, $\textbf{z}_{j}^v$ and  $\textbf{c}_{j}^v$ represent the interest embedding and social influence embedding of user $u_i$ and item $v_j$, respectively.

 To ensure independence between the decomposed interest and social influence embeddings, we introduce the mutual information loss function and minimize it,
\begin{equation}
\label{i_mi}
\begin{split}
L_{u}^{MI}& = \frac{1}{N^2}\sum_{i=1}^N\sum_{k=1,k\not=i}^N [logQ_{\theta_{u}}(\textbf{c}_{i}^u|\textbf{z}_{i}^u)-logQ_{\theta_{u}}(\textbf{c}_{k}^u|\textbf{z}_{i}^u))]\\
          & = \frac{1}{N}\sum_{i=1}^N[logQ_{\theta_{u}}(\textbf{c}_{i}^u|\textbf{z}_{i}^u)-\frac{1}{N}\sum_{k=1,k\not=i}^NlogQ_{\theta_{u}}(\textbf{c}_{k}^u|\textbf{z}_{i}^u)]\\
          & = \frac{1}{N}\sum_{i=1}^N[\frac{(\mu_i^u-\textbf{c}_{i}^u)^2}{e^{log({\delta_i^u}^2)}}-\frac{1}{N}\sum_{k=1,k\not=i}^N\frac{(\mu_i^u-\textbf{c}_{k}^u)^2}{e^{log({\delta_i^u}^2)}}];\\
L_{v}^{MI}& = \frac{1}{N^2}\sum_{j=1}^N\sum_{k=1,k\not=j}^N[logQ_{\theta_{v}}(\textbf{c}_{j}^v|\textbf{z}_{j}^v)-logQ_{\theta_{v}}(\textbf{c}_{k}^v|\textbf{z}_{j}^v)]\\
          & = \frac{1}{N}\sum_{j=1}^N[logQ_{\theta_{v}}(\textbf{c}_{j}^v|\textbf{z}_{j}^v)-\frac{1}{N}\sum_{k=1,k\not=j}^NlogQ_{\theta_{v}}(\textbf{c}_{k}^v|\textbf{z}_{j}^v)]\\
           & = \frac{1}{N}\sum_{j=1}^N[\frac{(\mu_j^v-\textbf{c}_{j}^v)^2}{e^{log({\delta_j^v}^2)}}-\frac{1}{N}\sum_{k=1,k\not=i}^N\frac{(\mu_j^v-\textbf{c}_{k}^v)^2}{e^{log({\delta_j^v}^2)}}],\\
\end{split}
\end{equation}

where $(\textbf{c}_{i}^u, \textbf{z}_{i}^u)$/$(\textbf{c}_{j}^v, \textbf{z}_{j}^v)$ is the positive pair and $(\textbf{c}_{k}^u, \textbf{z}_{i}^u)$/$(\textbf{c}_{k}^v, \textbf{z}_{j}^v)$ is the negative pair.

The total mutual information loss function and  variational approximation loss are as follows, 
\begin{equation}
\begin{split}
L^{MI}& = L_{u}^{MI}+L_{v}^{MI},\\
L^{LLD}& = -L_{u}^{LLD}-L_{v}^{LLD}.
\end{split}
\end{equation}
%The overall variational approximation loss is,
%\begin{equation}
%L^{LLD} = -L_{u}^{LLD}-L_{v}^{LLD}.
%\end{equation}

\subsection{Regulatory Decoder}
In traditional causal recommendation methods, social influence bias is often discarded.
However, in our method, we argue that social influence bias is individual-specific.
The impact of social influence bias can be detrimental when a user interacts with an item recommended by friends, deviating from their genuine interests. 
Conversely, it can be beneficial when a user engages with an item recommended by friends that is of high quality and aligns with his interests.
Blindly eliminating this bias may lead to suboptimal recommendation performance.
Therefore, we present a regulatory decoder designed to regulate social influence bias by reasonably incorporating positive social influence embeddings and mitigating the negative ones.
This decoder encompasses a dynamic weight calculation module and a fusion module. 
The weight calculation module dynamically computes the weight of social influence embeddings, while the fusion module integrates interest and social influence embeddings into the final user and item representations.

Generally, when users click on or purchase a product recommended by their friends that doesn't align with their interests, these users are less inclined to engage with similar items. 
Conversely, if the recommended product resonates with the users' genuine preferences, they are more likely to make subsequent purchases of similar products.
Therefore, for a user-item interaction pair $(u_i,v_j)$ where $v_j$ is recommended by friends of user $u_i$, we assume that if user $u_i$ interacts with another item $v_k$ that is similar to $v_j$, the main reason for $u_i$ clicking on item $v_j$ is that $u_i$ likes $v_j$.
In such cases, the impact of friends is beneficial, we should incorporate positive social influence embeddings.
Conversely, it might be attributed to the influence of herd mentality rather than the personal preference for the item, we need to mitigate the negative social influence embeddings. For the user-item interaction pair $(u_i,v_j)$, if user $u_i$'s friends have interacted with the item $v_j$, we consider $v_j$ to be recommended by $u_i$'s friends.

First, we calculate the maximum similarity between item $v_j$ and $v_k$,
\begin{equation}
s = max_{v_k\in V_h, k\not=j}f(v_j,v_k), 
\end{equation}
where $f$ is the similarity function, and $V_h=\{v_1,v_2,...,v_m\}$ represents the historical interaction records of user $u_i$.

The value of $s$ indicates the main reason why user $u_i$ clicks on the item $v_j$. 
A higher value indicates that the user consistently clicks on items similar to $v_j$. 
Multiple clicks mean that the user likes item $v_j$, implying a positive impact from the social influence bias.
Conversely, a small value represents that the user dislikes the item $v_j$, signifying a negative impact from the social influence bias.
We set the similarity threshold $s'$ as 0.5 and we could obtain the weight of social influence embeddings according to the similarity $s$,
\begin{equation}
\alpha = \begin{cases}
   1 , \quad s>s' \\
   0 , \quad s<=s'
\end{cases}
\end{equation}

It is worth noting that within the user-item interaction pair $(u_i,v_j)$, if the item $v_j$ is not recommended by the friends of user $u_i$, we assume that $u_i$ clicked on item $v_j$ based on personal preferences, entirely independent of any influence from the friends of user $u_i$. Consequently, we set the weight $\alpha$ to $0$.

Subsequently, a more accurate user representation $\textbf{h}_i^u$ and item representation $\textbf{h}_j^v$ can be learned as follows,
\begin{equation}
\begin{split}
\textbf{h}_i^u &= \textbf{z}_i^u + \alpha\textbf{c}_i^u,\\
\textbf{h}_j^v &= \textbf{z}_j^v + \alpha\textbf{c}_j^v.
\end{split}
\end{equation}

After reconstructing the user representation $\textbf{h}_i^u$ and item representation $\textbf{h}_j^v$, we concatenate them and put them into the final prediction layers,
\begin{equation}
\hat{y}_{ij} = MLP(\textbf{h}_i^u \oplus \textbf{h}_j^v).
\end{equation}

We aim to minimize the following loss function, 
% $L(r_{ij},\hat{r}_{ij})$, which represents the mean squared error in the prediction stage,

\begin{equation}
L^O=\frac{1}{N}\sum_{i,j=1}^N l(y_{ij}-\hat{y}_{ij}),
\end{equation}

where $N$ is the batch size, $l$ denotes the mean squared error in the rating prediction task and the cross-entropy loss function in the ranking task.

%\subsection{Model Optimization}
Finally, we optimize all parameters by minimizing the final loss function $L$,

\begin{equation}
L= L^O+\lambda (L^{MI}+L^{LLD}),
\label{eq:loss}
\end{equation}

where $\lambda$ is the weight parameter.

\section{Experiments}
In this section, we conduct a series of comprehensive experiments on four publicly available datasets to evaluate the performance of our proposed model CDRSB in the rating prediction task and ranking task. We aim to answer the following questions.
\begin{itemize}
\item RQ1: Does our model achieve superior performance compared to other state-of-the-art baseline methods?
\item RQ2: How do different components, such as mutual information-based objectives and social influence embeddings, affect the outcomes of our model?
\item RQ3: Are the interest and social influence embeddings we've acquired genuinely disentangled?
\item RQ4: How does our model's performance vary with different hyper-parameters?
\item RQ5: What is the primary reason for user interaction with an item?
 \end{itemize}
\subsection{Experimental Settings}
\subsubsection{Datasets}
We conduct experiments on four large-scale datasets: Ciao\footnote{http://www.cse.msu.edu/~tangjili/trust.html}, Epinions\footnotemark[\value{footnote}], Dianping\footnote{https://lihui.info/data/dianping/}, and Douban book\footnote{https://www.dropbox.com/s/u2ejjezjk08lz1o/Douban.tar.gz?e=1\&dl=0}.
These datasets comprise user-item ratings along with user trust relationships.
The Ciao and Epinions datasets are collected from popular social websites Ciao\footnote{http://www.ciao.co.uk} and Epinions\footnote{http://www.epinions.com}, where users have the ability to rate items and establish social connections by adding friends.
The Dianping dataset is collected from a leading local restaurant search and review platform in China\footnote{https://www.dianping.com}. This dataset is crawled by authors in \cite{li2015overlapping}. 
% The Yelp datasets is extracted from Yelp platform\footnote{https://www.yelp.com}, offering valuable insights into user reviews and business details.
The Douban book dataset is extracted from a Chinese book forum\footnote{https://book.douban.com/}.
The rating scale in all datasets ranges from 1 to 5. For the ranking task, we discretize the ratings into binary values of 0 and 1 to indicate whether the user has interacted with the item or not.
To ensure data quality and address sparsity issues, we apply filtering criteria to remove records with insufficient interactions.
% For the Dianping dataset, we remove samples with less than 10 interactions between users and items.
For the Epinions dataset, we delete records with fewer than five interactions.
For the Ciao dataset, we remove samples with less than three interactions between users and items.
For the Dianping and Douban book datasets, samples involving less than ten interactions between users and items are omitted.
Given the substantial data volume in both Dianping and Douban book datasets, we employ random sampling to enhance training efficiency.
Table \ref{tab2} provides an overview of the basic statistics of these datasets.

\begin{table*}[h]
\centering
\caption{Statistic of the datasets: Ciao, Epinions, Dianping and Douban book}.\label{tab2}
\begin{tabular}{c c c c c}
\hline
Datasets & Ciao & Epinions & Dianping & Douban book\\
\hline
\# Users & 7108 & 20461 &20000  &7000 \\
\# Items & 21978 & 31678 &9511  &16421 \\
\hline
\# Ratings & 184960 & 545861 &725637  & 443334 \\
Rating Density & 0.119\% & 0.084\% & 0.380\% & 0.386\%\\
\hline
\# Social Connections & 53019&  311235 &77146 & 11267\\
% Item Relation Connections &768& 864 \\
Social Connection Density & 0.105\% & 0.074\% & 0.019\% & 0.023\% \\
\hline
\end{tabular}

\end{table*}

\subsubsection{Evaluation Metrics}
To assess the recommendation performance of the CDRSB model and baselines, we depend on rating prediction and ranking metrics.

\textbf{Rating Prediction Metrics.} 
We utilize two commonly used evaluation metrics: root mean squared error (RMSE) and mean absolute error (MAE). 
These metrics are widely employed in collaborative prediction algorithms \cite{wang2015collaborative}. %[Collaborative deep learning for recommender systems],
RMSE measures the square root of the average squared difference between the predicted ratings $\hat{y}_i$ and the true ratings $y_i$ for a set of $N$ instances. A lower RMSE indicates better accuracy in predicting ratings.
MAE calculates the average absolute difference between the predicted ratings $\hat{y}_i$ and the true ratings $y_i$. 

\textbf{Ranking Metrics.}
We utilize two widely adopted metrics, Hit Rate (HR) and Normalized Discounted Cumulative Gain (NDCG), to assess the ranking performance. 
HR measures the proportion of samples where a user-interacted item appears within the top-K recommended items.
NDCG considers both the relevance and the position of the recommended items, assigning higher weights to items ranked higher in the list.

\subsubsection{Baseline Methods}
To verify the effectiveness of our model, we compare the performance of CDRSB with three sets of baseline methods: traditional recommendation systems, social recommendation systems, and causal recommendation systems.
We carefully selected several of the most representative methods from each category.
This comprehensive comparison allows us to assess the relative performance of CDRSB against different types of recommendation approaches and gain insights into its strengths and advantages.
% The summary of applied technologies in various baseline methods is presented in Table \ref{tab3}.
Furthermore, distinct sets of baselines are employed for the rating prediction and ranking tasks, as outlined in Table \ref{baseline summary}.
\begin{table}[htbp]
\centering
\caption{Summary of baselines for rating prediction and ranking tasks.}\label{baseline summary}
\begin{tabular}{c|c|c}
\hline
&Rating Prediction&Ranking\\
\hline
\multirow{2}*{\makecell{Traditional\\ recommendations}}&NeuMF&NeuMF\\
&PMF & LightGCN\\
\hline
\multirow{3}*{\makecell{Social \\recommendations}}&SocialMF&DiffNet\\
&GraphRec&DiffNet++\\
&ConsisRec&\\
\hline
\multirow{3}*{\makecell{Causal \\recommendations}}&CausE&DICE\\
&D2Rec&D2Rec\\
&IPS-MF &SIDR\\
\hline
\end{tabular}
\end{table}
\begin{itemize}
\item  \textbf{Traditional recommendation systems} 
%DCE-SR use auxiliary network information to learn the original user and item embeddings. 
%Therefore we compare DCE-SR with some traditional methods which don't use any side information to validate the effectiveness of network information.
\begin{itemize}
\item \textbf{PMF} \cite{mnih2007probabilistic} is a traditional recommendation method that decomposes the user-item rating matrix into low-dimensional latent feature matrices. It learns the relationships among these latent features for rating prediction.
\item \textbf{NeuMF} \cite{he2017neural} combines collaborative filtering and neural network techniques to capture user-item interactions and make accurate predictions. 
For the rating prediction task, we modify its loss function to the square loss.
% It focuses on implicit feedback, but we modify its loss function to the square loss for rating prediction to align with our model's emphasis on explicit feedback.
\item \textbf{LightGCN} \cite{he2020lightgcn} is a simple GCN model that directly propagates user and item embeddings through the user-item interaction graph without introducing complex operations or auxiliary information.
\end{itemize}
\item \textbf{Social recommendation systems}
% Compared to these social recommendation baseline methods, CDLSB considers the social influence bias in social recommendations.
\begin{itemize}
\item \textbf{SocialMF} \cite{jamali2010matrix}
incorporates user trust information into a matrix factorization model, leveraging user social relationships to infer user ratings for items.
 \item \textbf{GraphRec} \cite{fan2019graph}
proposes a framework for social recommendation that leverages Graph Neural Networks to learn user and item representations with user-item interactions and user social networks.
 \item \textbf{ConsisRec} \cite{yang2021consisrec}
is an enhanced method based on GraphRec that effectively addresses the issue of social inconsistency by leveraging consistent neighbor aggregation.
 \item \textbf{DiffNet} \cite{wu2019neural} proposes a deep influence propagation model to simulate how users are influenced by the recursive social diffusion process for social recommendation.
 \item \textbf{DiffNet++} \cite{wu2020diffnet++} is an improved model based on DiffNet, incorporating the modeling of interest diffusion with a user-item graph.
\end{itemize}
\item \textbf{Causal recommendation systems}
\begin{itemize}
\item \textbf{IPS-MF} \cite{liang2016causal} utilizes inverse propensity score (IPS) to mitigate the selection bias produced by exposure data.
\item \textbf{CausE} \cite{bonner2018causal} proposes a domain adaptation method that trains the model by utilizing biased data and predicts recommendation results based on random exposure. 
\item \textbf{DICE} \cite{zheng2021disentangling} first disentangles user and item embeddings into interest and conformity with cause-specific data and then eliminates the confounding effect of popularity bias.
\item \textbf{D2Rec} \cite{sheth2022causal} disentangles user and item representations into inherent,
confounder, and exposure factors, and then mitigates social influence bias by a reweighting function.
\item \textbf{SIDR} \cite{sheth2023causal} causally disentangles the user and item latent features to mitigate social influence bias in implicit feedback for social recommendation.
\end{itemize}
\end{itemize}
\subsubsection{Parameter Settings}
We implement the CDRSB model using Python with the Pytorch framework, all baseline methods are conducted based on their GitHub source code and carefully adjusted the hyperparameters.
We randomly split datasets into training, test, and validation sets according to an 8:1:1 ratio. The optimal hyperparameters are obtained by optimizing the loss function \eqref{eq:loss} using the RMSprop optimizer.
Based on the validation set, we evaluate the performance of the model using different parameter combinations. The embedding size of original embedding, decomposing embedding, and batch size are searched within the range of [8, 16, 32, 64, 128, 256]. Ultimately, we set them to 64, 64, and 128, respectively.
We conduct tests with different learning rates [0.0001, 0.0005, 0.001, 0.005, 0.01, 0.05, 0.1] and $\lambda$ values [0.0001, 0.001, 0.01, 0.1]. We determine that the optimal learning rate is 0.0001, and we set $\lambda$ to 0.001.
% We adopt two hidden layers for each neural component in the overall framework.
To prevent overfitting, we apply batch normalization, dropout, and early stopping techniques where the training is stopped when the test evaluation metrics increase for 5 epochs.
\subsection{Results and Analysis (RQ1)}
We evaluate the performance of CDRSB and the baselines using commonly used evaluation metrics for rating prediction task w.r.t. RMSE and MAE and ranking task w.r.t. HR@10 and NDCG@10 on four datasets. The results for each task are shown in Table \ref{comparison_rating} and Table \ref{comparison_ranking}.
\begin{table*}[!htb]
\centering
\small
\caption{Overall performance comparison for the rating prediction task. The optimal performance is highlighted using bold fonts, and the second-best performance is denoted by underlines.}\label{comparison_rating}
\begin{tabular}{c|c|c c|c c|c c|c c}
\hline
  \multirow{2}*{Method type}& \multirow{2}*{Method} & \multicolumn{2}{c|}{Epinions} &  \multicolumn{2}{c|}{Ciao} & \multicolumn{2}{c|}{Dianping}  &  \multicolumn{2}{c}{Douban book}\\
\cline{3-10}
& & RMSE & MAE & RMSE &MAE& RMSE &MAE& RMSE&MAE\\
\hline
 \multirow{2}*{\makecell{Traditional\\ recommendations}}
 & PMF&1.2905 & 1.0203 & 1.1309&0.9107&1.0260&0.8530&1.0176&0.8425\\
  &NeuMF&1.1290 & 0.9040 & 1.0713&0.8145&0.9606&0.7887&0.9513&0.7802\\
 \hline
 \multirow{3}*{\makecell{Social \\recommendations}}&SociaMF & 1.0934 & 0.8442&1.0534&0.8223&0.9512&0.7945&0.9428&0.7845\\
&GraphRec& 1.0657& 0.8134& 0.9978&  0.7523&0.8205&0.6023&0.8116 &0.5928\\
&ConsisRec &1.0467& 0.8096&0.9823&0.7436&0.8073&0.5796&0.8025&0.5768\\
\hline
 \multirow{4}*{\makecell{Causal \\recommendations}}
 &IPS-MF &1.1023  &0.8813 &1.0423 &0.7904&0.9649 &0.7514 & 0.9584&0.7461\\
 &CausE &1.1145  &0.8956 &1.0378 &0.7810&0.8546 &0.6481 & 0.8475&0.6362\\
 &D2Rec & \underline{1.0223} & \underline{0.8545}&\underline{0.9394}&\underline{0.7103}&\underline{0.7256}&\underline{0.5478}&\underline{0.7124}&\underline{0.5405}\\
&\textbf{CDRSB} &\textbf{0.9404} & \textbf{0.7078} & \textbf{0.8246}&\textbf{0.5950}& \textbf{0.6759}&\textbf{0.5128}& \textbf{0.6695}&\textbf{0.5277}\\
\hline
&Imp.\%& $\uparrow$ 8.19\%&$\uparrow$ 14.67\%&$\uparrow$ 11.48\%&$\uparrow$ 11.53\%&$\uparrow$ 4.97\%&$\uparrow$ 3.50\%&$\uparrow$ 4.29\%&$\uparrow$ 1.28\%\\
\hline
\end{tabular}
\end{table*}

\begin{table*}[!htb]
\centering
\small
\caption{Overall performance comparison for ranking task. The optimal performance is highlighted using bold fonts, and the second-best performance is denoted by underlines.}\label{comparison_ranking}
\begin{tabular}{c|c|c c|c c|c c|c c}
\hline
  \multirow{2}*{Method type}& \multirow{2}*{Method} & \multicolumn{2}{c|}{Epinions} &  \multicolumn{2}{c|}{Ciao}&  \multicolumn{2}{c|}{Dianping}&  \multicolumn{2}{c}{Douban book}\\
\cline{3-10} 
& & HR@10 & NDCG@10 & HR@10 &NDCG@10& HR@10 &NDCG@10& HR@10 &NDCG@10\\
\hline
 \multirow{2}*{\makecell{Traditional \\recommendations}}
&NeuMF&0.3815 & 0.2389 & 0.3393&0.2231&0.4123&0.2701&0.4163&0.2572\\
& LightGCN&0.4013 & 0.2527 & 0.3456&0.2367&0.4234&0.2825&0.4389&0.2826\\
 \hline
 \multirow{2}*{\makecell{Social \\recommendations}}&DiffNet & 0.4264 & 0.2756&0.3589&0.2501&0.4495&0.3026&0.4527&0.3009\\
&DiffNet++& 0.4479& 0.3016& 0.3670&  0.2704&0.4726&0.3313&0.4848&0.3428\\
\hline
 \multirow{4}*{\makecell{Causal \\recommendations}}
 &DICE &0.6735  &0.3902 &0.5572 &0.3301&0.6972&0.4231&0.7026&0.4237\\
 &D2Rec & 0.6703 &0.3876 &0.5674 & 0.3356&0.7004&0.4294&0.7128&0.4162\\
 &SIDR &\underline{0.6824}  &\underline{0.4036} &\underline{0.5863} &\underline{0.3528}&\underline{0.7183}&\underline{0.4368}&\underline{0.7345}&\underline{0.4596}\\
 
&\textbf{CDRSB} &\textbf{0.8082} & \textbf{0.5434} & \textbf{0.7599}&\textbf{0.4772}& \textbf{0.7768}&\textbf{0.5146}& \textbf{0.8260}&\textbf{0.5791}\\
\hline
&Imp.\%& $\uparrow$ 12.58\%&$\uparrow$ 13.98\%&$\uparrow$ 17.36\%&$\uparrow$ 12.44\%&$\uparrow$ 5.85\%&$\uparrow$ 7.78\%&$\uparrow$ 9.15\%&$\uparrow$ 11.95\%\\
\hline
\end{tabular}
\end{table*}

\begin{itemize}
\item 
% Our model, CDRSB, outperforms other baselines. 
% In terms of metrics for rating prediction (RMSE and MAE) and ranking (HR@10 and NDCG@10) on two datasets (Ciao and Epinions), CDRSB demonstrates superior performance with average improvements of up to 20\% and 17.37\% as well as 28.46\% and 15.56\% over the best baseline models NeuMF and LightGCN in traditional recommendation methods,  11.43\% and 9.11\% as well as 25.07\% and 11.43\% over the best baseline ConsisRec and DiffNet++ in social recommendation methods, 8.07\% and 9.69\% as well as 2.38\% and 2.20\% over the best baseline D2Rec and SIDR in causal recommendation methods.
% The results indicate that the effectiveness and rationality of CDRSB. 
Our model, CDRSB, outperforms other baselines in terms of metrics for rating prediction (RMSE and MAE) and ranking (HR@10 and NDCG@10) on four datasets.
CDRSB demonstrates superior performance, achieving average improvements of up to 39.04\% and 26.49\% in terms of HR@10 and NDCG@10 over the best baseline model LightGCN in traditional recommendation methods. Furthermore, it outperforms the best baseline model DiffNet++ by an average of 34.97\% and 21.71\% in terms of HR@10 and NDCG@10 in social recommendation methods. Additionally, CDRSB exhibits better performance than the best baseline model D2Rec, with improvements of 7.23\% and 7.75\% in terms of RMSE and MAE in causal recommendation methods. The results indicate the effectiveness and rationality of CDRSB.

\item Compared to D2Rec and SIDR, which mitigate social influence bias, CDRSB achieves the highest performance. This demonstrates that properly regulating social influence bias could improve the model's overall performance.
\item Social recommendation methods based on causal debiasing, such as D2Rec and SIDR, outperform traditional social recommendation methods across all evaluation metrics. This emphasizes the significance of mitigating social influence bias in enhancing recommendation performance.
\item SocialMF, GraphRec, and ConsisRec, which incorporate social network information, outperform traditional recommendation models NeuMF and PMF. This is attributed to the fact that social networks can complement user preferences, particularly when user features are sparse.
\item Both NeuMF and PMF utilize rating information for recommendations. However, NeuMF outperforms PMF, indicating the superior learning capability of deep learning models in recommendation systems.
\item GraphRec outperforms SocialMF by an average of 8.63\% and 12.12\% in terms of RMSE and MAE across four datasets.
These results highlight the advantages of GNN and the incorporation of rating information into the learning process of user and item embeddings.
\end{itemize}
\subsection{Ablation Studies (RQ2)}
To evaluate the effectiveness of each component in CDRSB, we conduct ablation experiments on four datasets.
We create three variants of CDRSB, denoted as w/o wt, w/o sl, and w/o mi by removing specific components.
\begin{itemize}
\item w/o wt: It eliminates the weight calculation module, thereby fixing the weights of social influence embeddings at 1.
\item w/o sl: It removes user and item social influence embeddings and only uses user and item interest embeddings for recommendation.
%\item w/o int: It deletes user and item interest embeddings and only utilizes user and item social influence embeddings for recommendation.
% \item Variant D: It eliminates the GNN-based learning network. Instead, it directly utilizes user ID and item ID to learn original user and item embeddings.
\item w/o mi: It eliminates the mutual information minimization objective from the joint loss.
\end{itemize}
The results of the ablation studies for rating prediction and ranking tasks are presented in Table \ref{ablation_study_rating} and Table \ref{ablation_study_ranking}.

\begin{table*}[!htb]
\centering
\caption{Results of the ablation studies on the rating prediction task.}
\begin{tabular}{c|cc|c c|c c|c c}
\hline
 &\multicolumn{2}{c|}{ Ciao} & \multicolumn{2}{c|}{Epinions}& \multicolumn{2}{c|}{Dianping}& \multicolumn{2}{c}{Douban book}\\
\hline
Method & RMSE & MAE& RMSE & MAE& RMSE & MAE& RMSE & MAE\\
\hline
w/o wt & 0.8379 &0.6126&0.9605&0.7274&0.6934&0.5326&0.6836&0.5529\\
w/o sl & 0.8345 &0.6069&0.9592&0.7201&0.6883&0.5278&0.6796&0.5401 \\
% D &1.0435  &0.7795&1.0778&0.8123& 0.8745&0.6902 \\
w/o mi &0.8308  &0.6006&0.9547&0.7156&0.6825&0.5234&0.6754&0.5357\\
CDRSB& \textbf{0.8246}&\textbf{0.5950}&\textbf{0.9404}&\textbf{0.7078}&\textbf{0.6759}&\textbf{0.5128}&\textbf{0.6695}&\textbf{0.5277}\\
\hline
\end{tabular}
\label{ablation_study_rating}
\end{table*}

\begin{table*}[!htb]
\centering
\caption{Results of the ablation studies on the ranking task.}
\begin{tabular}{c|cc|c c|c c|c c}
\hline
 &\multicolumn{2}{c|}{ Ciao} & \multicolumn{2}{c|}{Epinions} & \multicolumn{2}{c|}{Dianping} & \multicolumn{2}{c}{Douban book}\\
\hline
Method & HR@10 & NDCG@10& HR@10 & NDCG@10 & HR@10 & NDCG@10 & HR@10 & NDCG@10\\
\hline
w/o wt & 0.7302&0.4501&0.7598&0.4938&0.7310&0.4709&0.7902&0.5313\\
w/o sl & 0.7268&0.4439& 0.7603&0.4967&0.7278&0.4625&0.7810&0.5148 \\
% D &1.0435  &0.7795&1.0778&0.8123& 0.8745&0.6902 \\
w/o mi & 0.7325 &0.4503&0.7663&0.5015&0.7346&0.4739&0.7929&0.5393\\
CDRSB& \textbf{0.7599}&\textbf{0.4772}&\textbf{0.8082}&\textbf{0.5434}&\textbf{0.7768}&\textbf{0.5146}&\textbf{0.8260}&\textbf{0.5791}\\
\hline
\end{tabular}
\label{ablation_study_ranking}
\end{table*}

Based on the above results, we can observe that each component of the overall model plays a crucial role.
% without the GNN-based learning network, the model D performs the poorest, which shows the importance of network information.
The model w/o mi, which eliminates mutual information restraints, experiences a significant drop in performance.
This outcome might be attributed to the fact that mutual information minimization ensures that the decoupled interest embedding and social influence embedding have non-redundant information, leading to enhanced performance.
Comparing CDRSB with w/o wt, where negative social influence embeddings are incorporated, it becomes evident that the weight calculation module plays a pivotal role in effectively controlling social influence bias.
Moreover, the inferior performance of model w/o sl emphasizes the significant contributions of the positive social influence embeddings to the outcome.

\subsection{Visualization of Disentangled Embeddings (RQ3)}
\begin{figure}[!htb]
\centering
\subfloat[Initialization]{\includegraphics[width=1.6in]{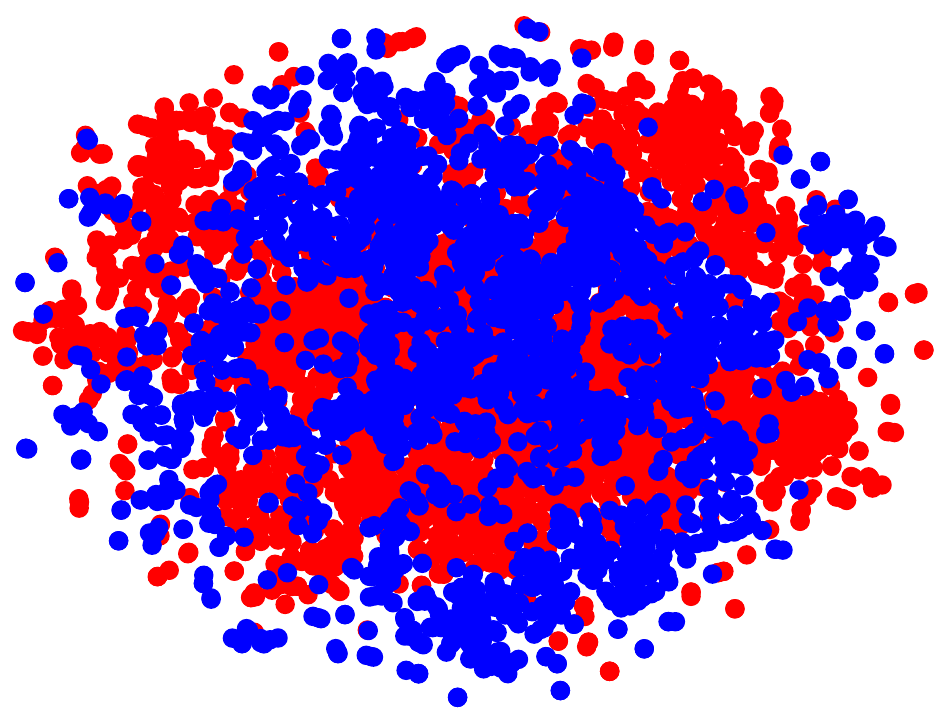}
}
\hfil
\subfloat[Convergence]{\includegraphics[width=1.6in]{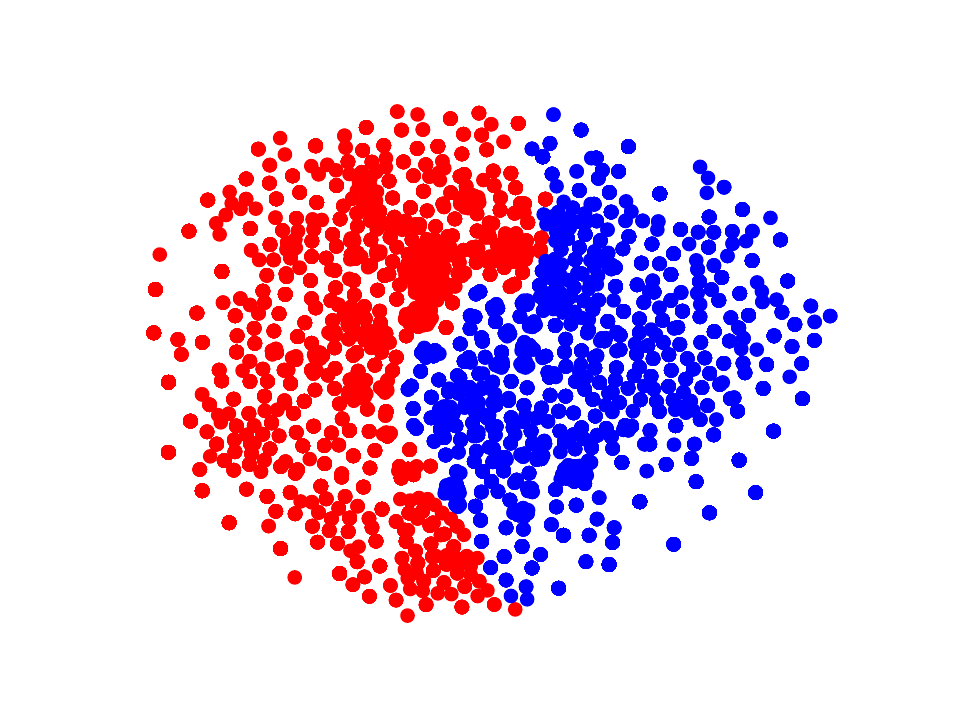}
}
\hfil
\caption{Visualization of user's interest embedding (red points) and social influence embedding (blue points) for different stages: (a) initialization, (b) convergence on the dataset Epinions.}
\label{user_emb}
\end{figure}

\begin{figure}[!htb]
\centering
\subfloat[Initialization]{\includegraphics[width=1.6in]{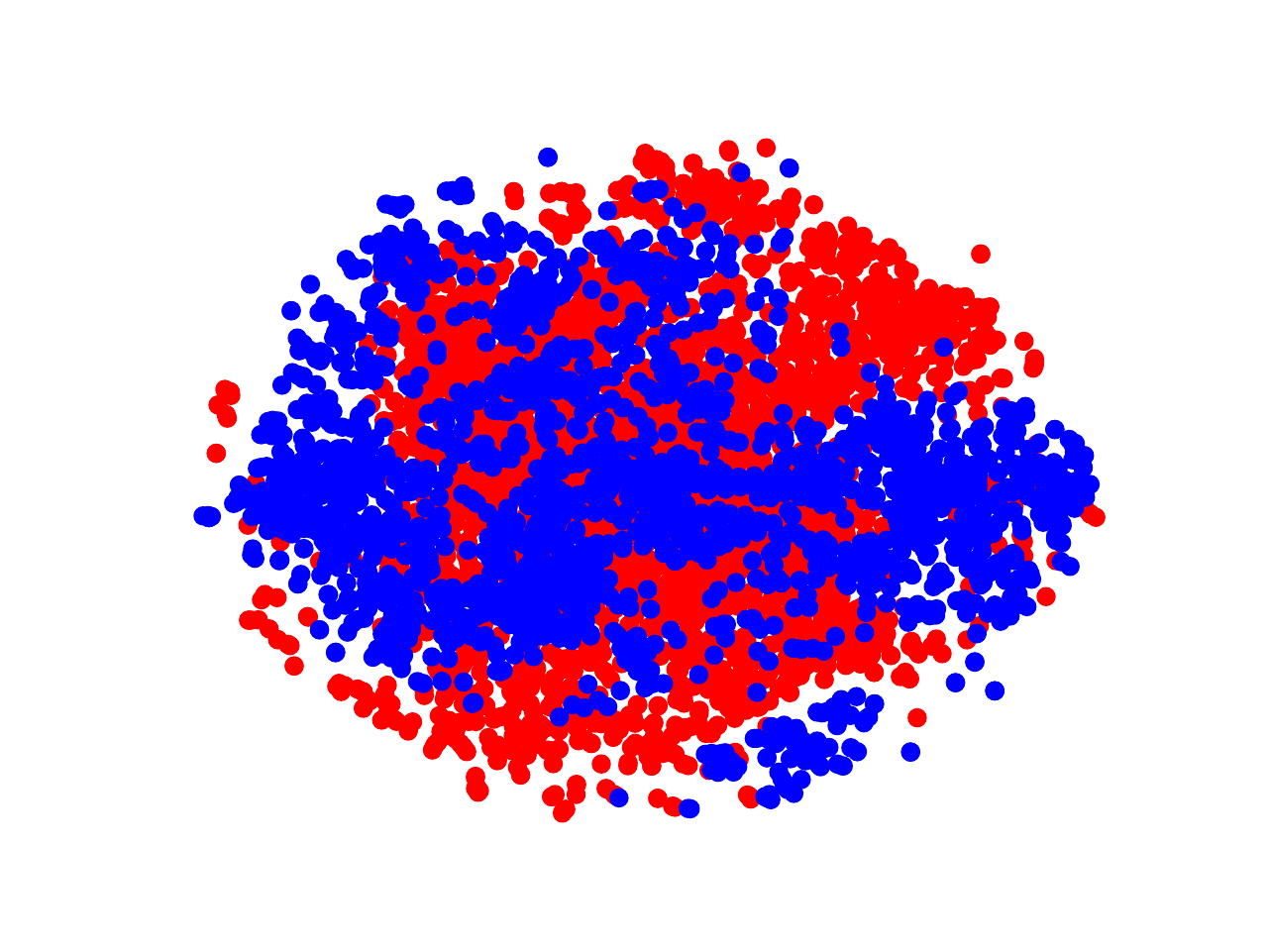}
}
\hfil
\subfloat[Convergence]{\includegraphics[width=1.6in]{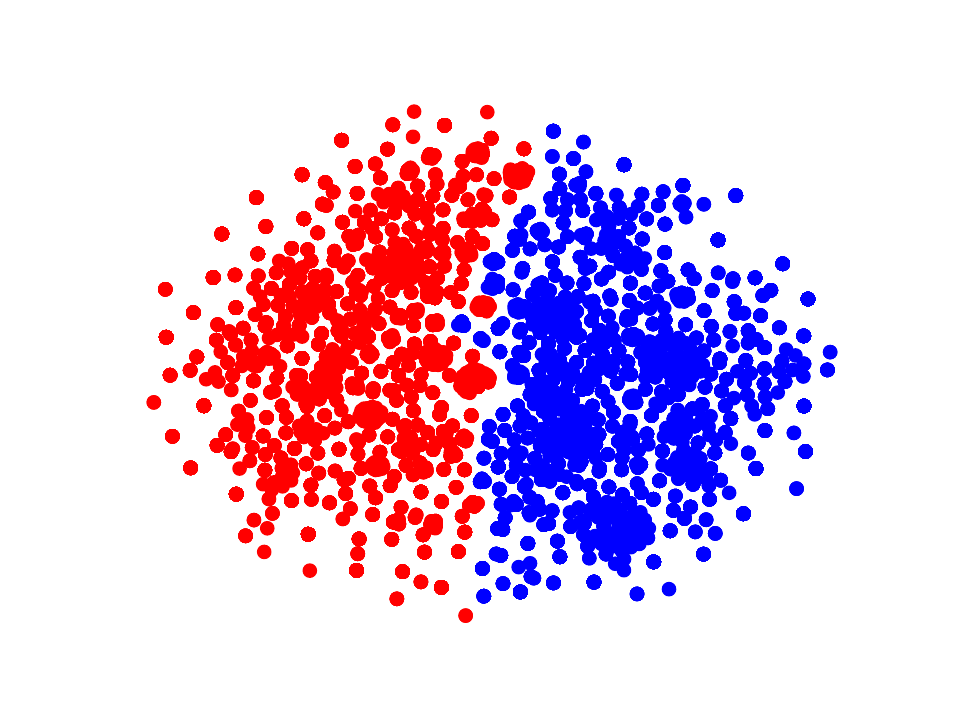}
}
\hfil
\caption{Visualization of item's interest embedding (red points) and social influence embedding (blue points) for different stages: (a) initialization, (b) convergence on the dataset Epinions.}
\label{item_emb}
\end{figure}

In this section, we aim to gain deeper insights into the impact of the disentangled encoder on the representative learning process in CDRSB.
Specifically, we investigate whether the interest embedding and social influence embedding become independent of each other during the training process.
We employ t-SNE \cite{van2008visualizing}, a data visualization technique that projects high-dimensional data into a lower-dimensional space, to visualize this phenomenon.
The disentangled interest embedding and social influence embedding for the rating prediction task on the Epinions dataset are visualized in Figure \ref{user_emb} and Figure \ref{item_emb}.

We observe that with an increasing number of model training iterations, a distinction emerges between the interest embedding represented in red and the social influence embedding represented in blue.
This clear separation validates the effectiveness of the disentangled encoder in disentangling and distinguishing the interest and social influence within the embeddings.

\subsection{Parameter Sensitivity (RQ4)}
In this section, we evaluate the performance of CDRSB under various settings of two crucial parameters: the decoupling embedding dimension and the weight parameter $\lambda$.
\begin{itemize}
    \item \textbf{The impact of decoupling embedding dimension.} For the rating prediction and ranking tasks, we present the comparative results in Figure \ref{emb_dim_rating} and Figure \ref{emb_dim_ranking}, respectively. It is observed that optimal performance is achieved when the dimension is set to 64.
Increasing the dimension improves the model's effectiveness. However, it is important to note that too large embedding dimensions may lead to overfitting. Thus, choosing an appropriate decoupled embedding dimension is crucial to balance model complexity and performance.
\begin{figure}[!htb]
\centering
\subfloat[]{\includegraphics[width=1.6in]{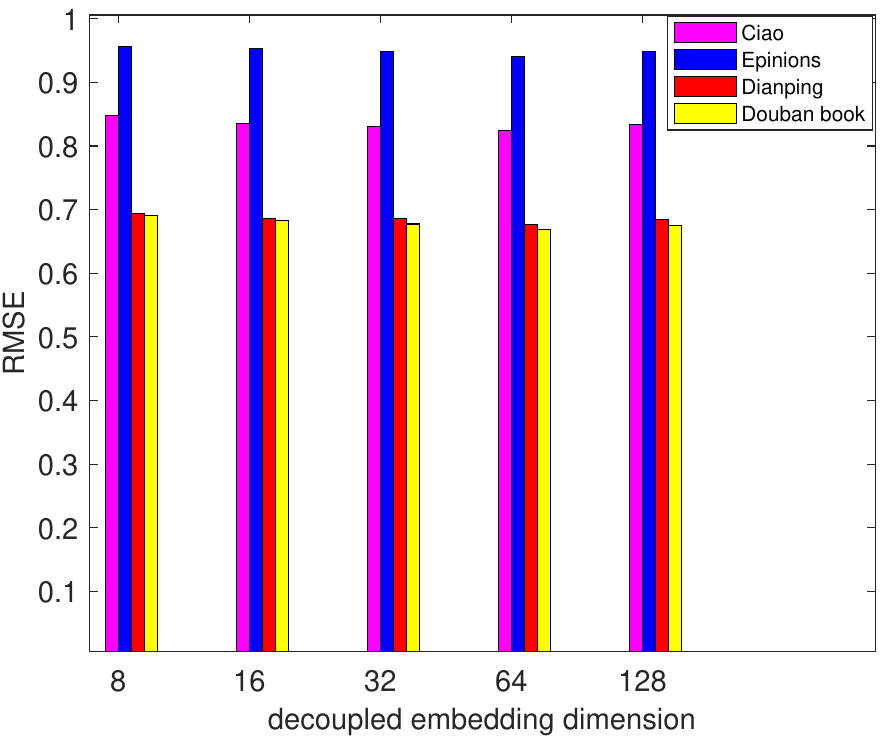}
\label{emb_dim_rmse}}
\hfil
\subfloat[]{\includegraphics[width=1.6in]{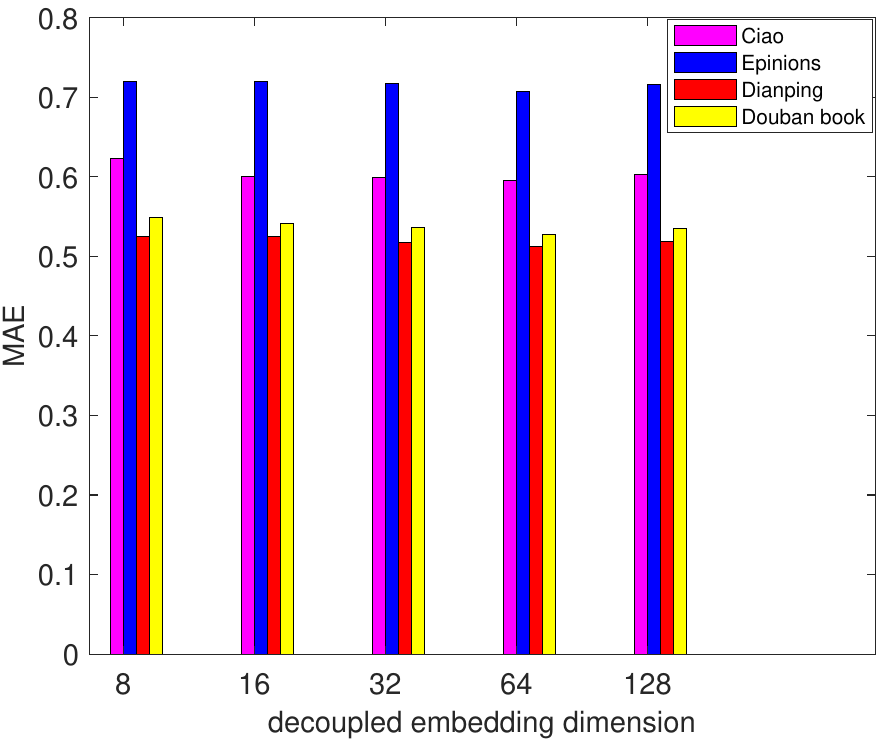}
\label{emb_dim_mae}}
\hfil
\caption{RMSE and MAE on different decoupled embedding dimensions: (a) RMSE  (b) MAE.}
\label{emb_dim_rating}
\end{figure}

\item \textbf{The impact of weight parameter $\lambda$.} The comparative results for rating prediction and ranking tasks are shown in Figure \ref{lambda_rating} and Figure \ref{lambda_ranking}. It can be observed that CDRSB obtains the optimal result when $\lambda$ is set to 0.001.
This indicates that a moderate weight parameter strikes a balance between incorporating the mutual information minimization objective and preserving the overall recommendation performance.

\begin{figure}[!htb]
\centering
\subfloat[]{\includegraphics[width=1.6in]{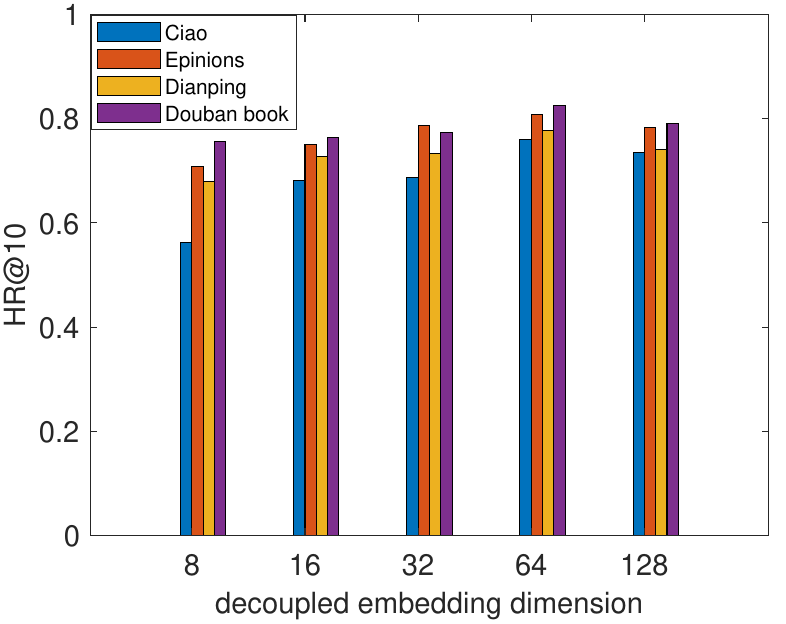}
\label{emb_dim_hr}}
\hfil
\subfloat[]{\includegraphics[width=1.6in]{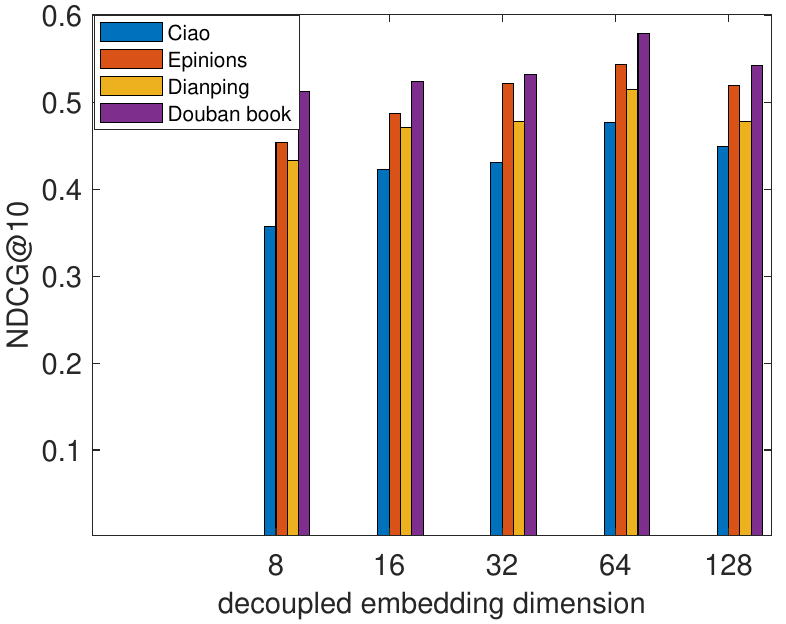}
\label{emb_dim_ndcg}}
\hfil
\caption{HR@10 and NDCG@10 on different decoupled embedding dimensions: (a) HR@10  (b) NDCG@10.}
\label{emb_dim_ranking}
\end{figure}

\end{itemize}

\begin{figure}[!htb]
\centering
\subfloat[]{\includegraphics[width=1.6in]{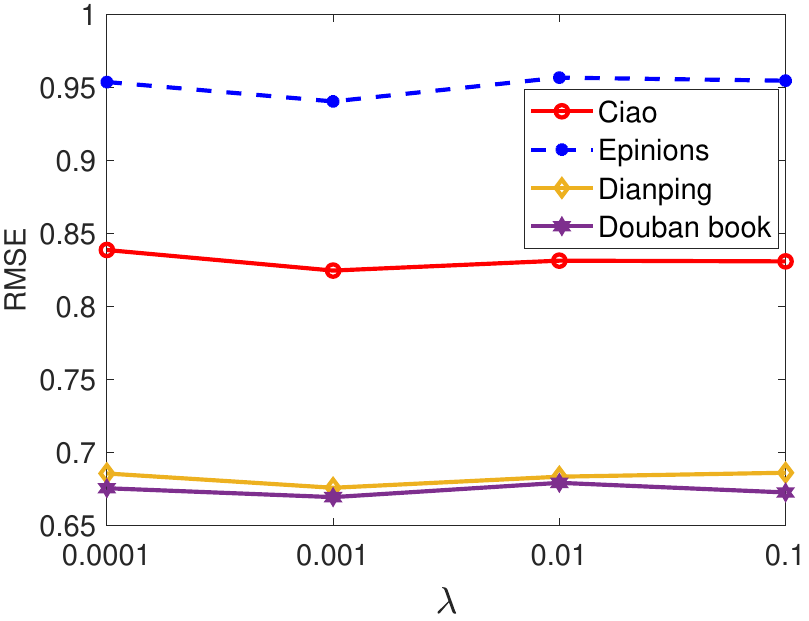}
\label{lambda_rmse.pdf}}
\hfil
\subfloat[]{\includegraphics[width=1.6in]{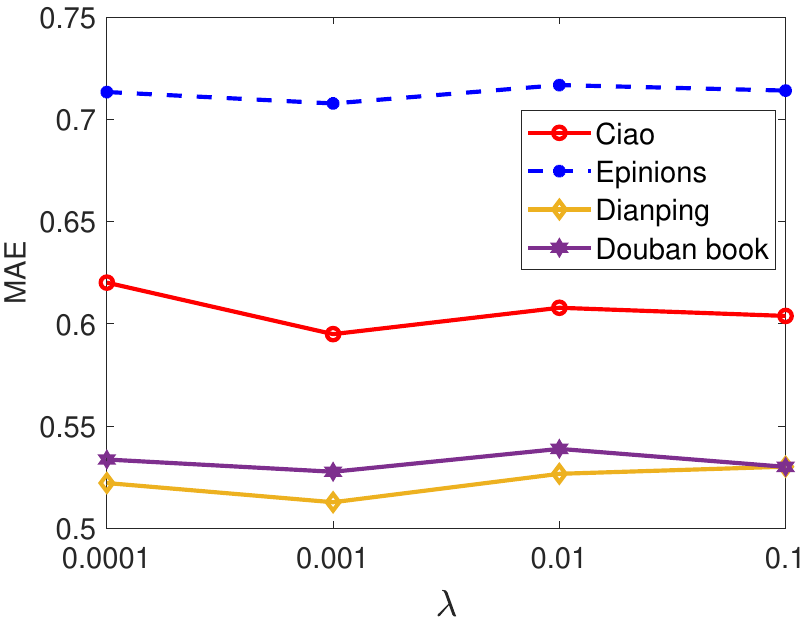}
\label{lambda_mae.pdf}}
\hfil
\caption{RMSE and MAE on different $\lambda$: (a) RMSE  (b) MAE.}
\label{lambda_rating}
\end{figure}

\begin{figure}[!htb]
\centering
\subfloat[]{\includegraphics[width=1.6in]{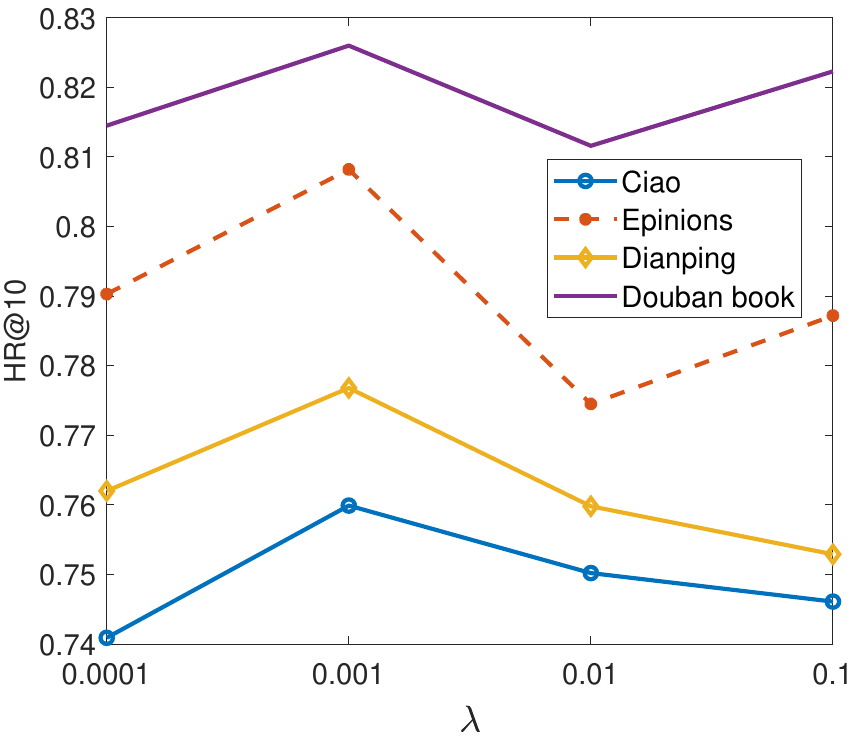}
\label{lambda_hrr}}
\hfil
\subfloat[]{\includegraphics[width=1.6in]{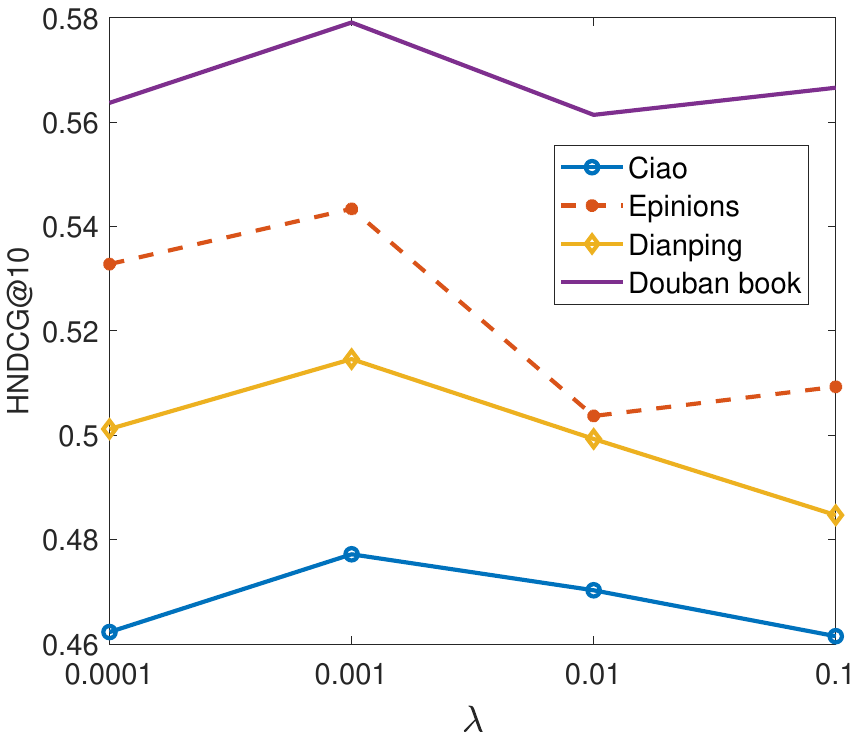}
\label{lambda_ndcg}}
\hfil
\caption{HR@10 and NDCG@10 on different $\lambda$: (a) HR@10  (b) NDCG@10.}
\label{lambda_ranking}
\end{figure}

\subsection{Case Study (RQ5)}
\begin{figure}[!htb]
\centering
\subfloat[]{\includegraphics[width=1.6in]{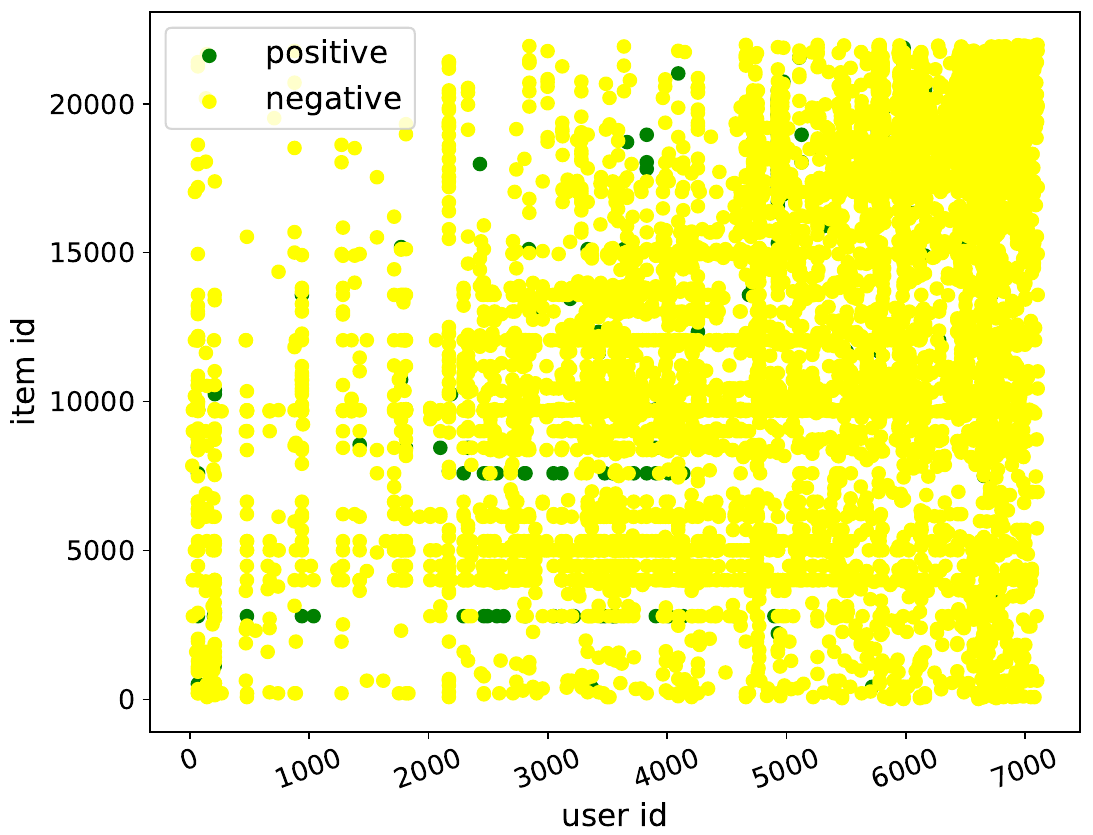}
\label{ciao_inter_reason}}
\hfil
\subfloat[]{\includegraphics[width=1.6in]{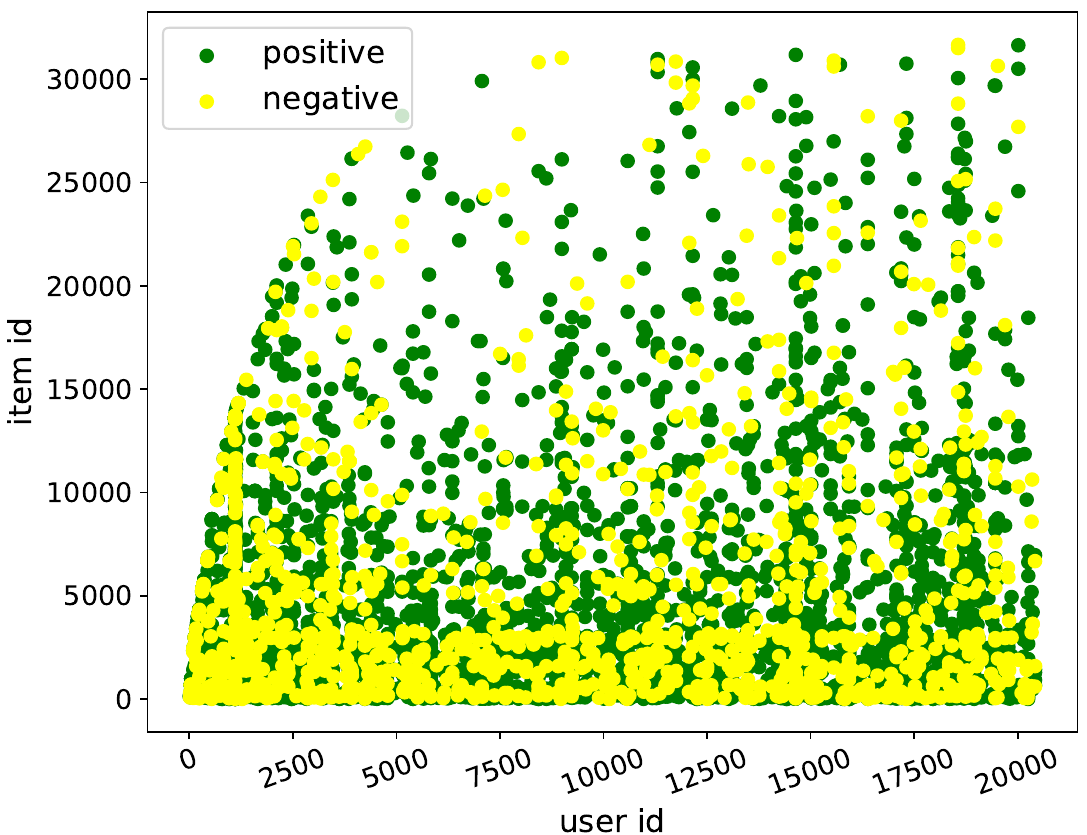}
\label{epinion_inter_reason}}
\hfil
\caption{The main reason for user-item interactions: (a) Ciao  (b) Epinions.}
\label{inter_reason_ciao_epinion}
\end{figure}

\begin{figure}[!htb]
\centering
\subfloat[]{\includegraphics[width=1.6in]{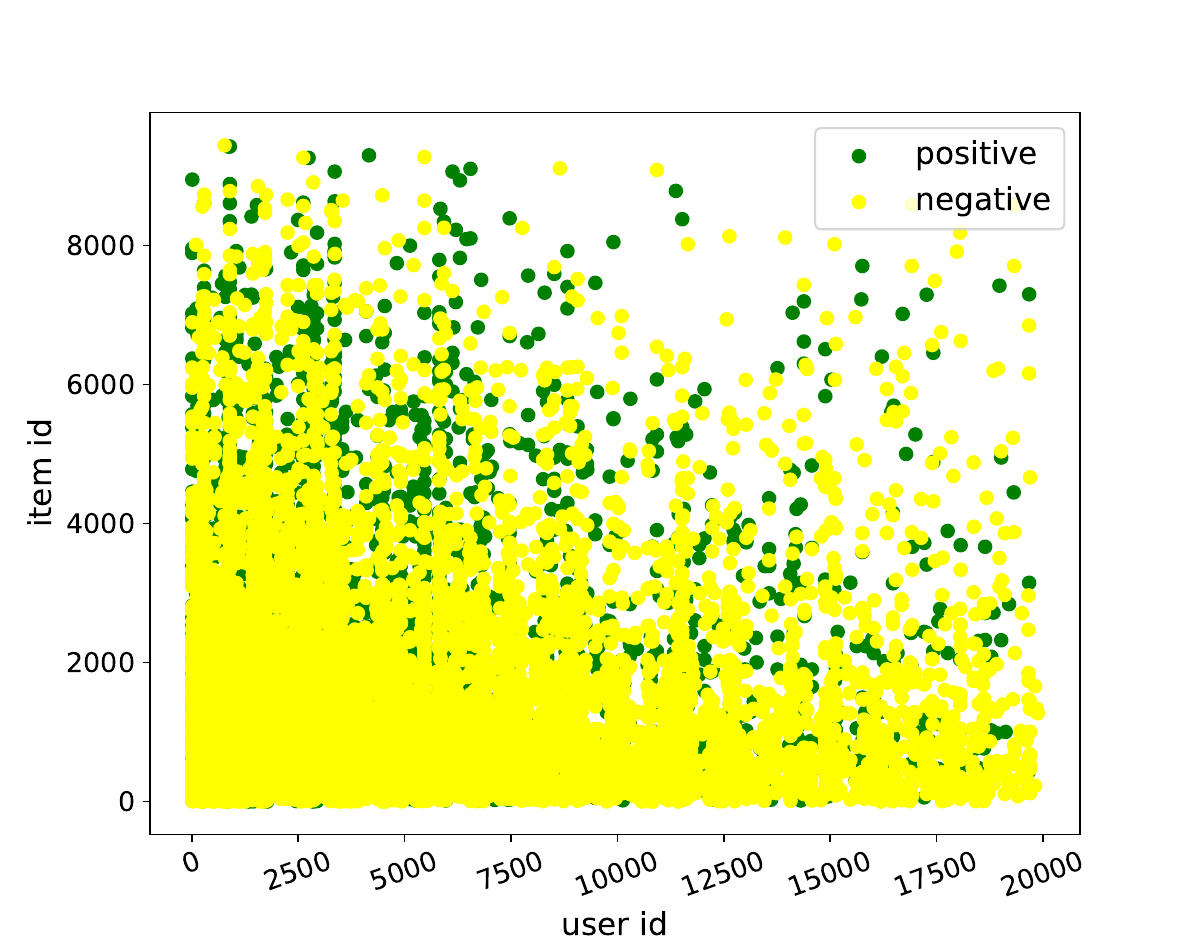}
\label{dianping_inter_reason}}
\hfil
\subfloat[]{\includegraphics[width=1.6in]{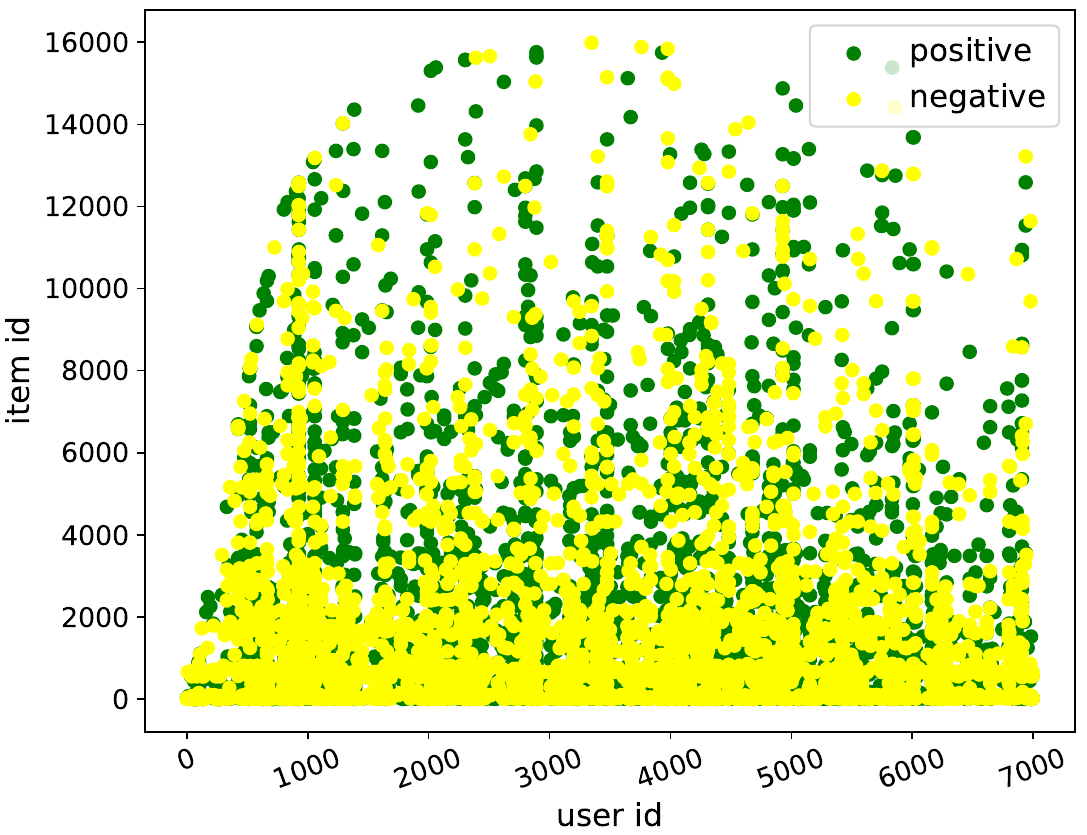}
\label{yelp_inter_reason}}
\hfil
\caption{The main reason for user-item interactions: (a) Dianping  (b) Douban book.}
\label{inter_reason_dianping_douban_book}
\end{figure}

In this section, we delve into the motivations behind user-item interactions, exploring whether they primarily stem from individual preferences or are influenced by herd mentality.
We randomly sample 1000 users who received recommendations from friends across all datasets and subsequently visualize the different effects of social influence embeddings on each user-item interaction pair for the rating prediction task, as shown in Figures \ref{inter_reason_ciao_epinion} and \ref{inter_reason_dianping_douban_book}.

In the Figures \ref{inter_reason_ciao_epinion} and \ref{inter_reason_dianping_douban_book}, the green points denote that the social embeddings have a positive effect, implying that recommendations from friends align with users' interests, showcasing interactions driven by personal preferences. Conversely, the yellow points signify a negative effect of social influence embeddings, indicating recommendations from friends that diverge from users' interests and reflect interactions influenced by herd mentality.

\section{Conclusion and Future Work}
% In this paper, we first explain the generation process of social influence bias in social recommendations from the causal perspective.
In this paper, we propose a causal disentanglement-based framework for regulating social influence
bias in social recommendation, named CDRSB.
First, due to the existence of the social network confounder, there are two paths between user and item embeddings and user-item ratings in social recommendations: a non-causal social influence path and a causal interest path.
Therefore, we propose a disentangled encoder that decomposes user and item embeddings into interest and social influence embeddings.
We leverage mutual information minimization techniques to ensure the independence of these two components from each other.
Next, we present a regulatory decoder to regulate social influence bias and integrate interest and social influence embeddings into more accurate user and item representations.
Experimental results on four datasets Ciao, Epinions, Dianping, and Douban book demonstrate the effectiveness of our proposed model CDRSB and its various components.

In the future, we plan to explore the following three directions. 
Firstly, we will investigate other effective disentanglement methods that can enhance the separation of user preferences and social influence bias. 
Secondly, we aim to incorporate additional side information, such as user search data or review texts, to assist in decomposing true preferences and social influence more accurately. 
Additionally, we intend to further disentangle users' interests into finer-grained factors, such as clothing color, style, and price.
\\ 

\noindent \textbf{CRediT authorship contribution statement}\\

\textbf{Li Wang}: Conceptualization, Investigation, Methodology,
Software, Writing - original draft, Writing - review \& editing.
\textbf{Min Xu}: Supervision, Writing - review \& editing.
\textbf{Quangui Zhang}: Writing - review \& editing.
\textbf{Yunxiao Shi}: Data curation.
\textbf{Qiang Wu}: Supervision, Writing - review \& editing.
%% If you have bibdatabase file and want bibtex to generate the
%% bibitems, please use
%%
 \bibliographystyle{elsarticle-num} 
 \bibliography{references}

%% else use the following coding to input the bibitems directly in the
%% TeX file.

% \begin{thebibliography}{00}

% %% \bibitem{label}
% %% Text of bibliographic item

% \bibitem{}

% \end{thebibliography}

\end{document}